\documentclass[a4paper,12pt]{article}

\usepackage{amsmath,amssymb,array,amsfonts}
\usepackage[dvipdfmx]{graphicx}
\usepackage{comment,color}
\usepackage{geometry}
\usepackage[bf, small, hang]{caption}
\usepackage{cite}

\allowdisplaybreaks

\def\e{{\rm e}}	
\def\bs{\boldsymbol}

\def\dis{\displaystyle}
\def\lra{\longrightarrow}
\newcommand{\be}{\begin{equation}}
\newcommand{\ee}{\end{equation}}
\newcommand{\bea}{\begin{eqnarray}}
\newcommand{\eea}{\end{eqnarray}} 

\newcommand{\sls}[1]{{\ooalign{\hfil/\hfil\crcr$#1$}}}

\makeatletter
	
	\@addtoreset{equation}{section}
\makeatother

\usepackage
[colorlinks=true,
 linkcolor=magenta,
 urlcolor=magenta,
 citecolor=cyan]
 {hyperref}

\renewcommand{\figureautorefname}{figure}

\begin{document}
\setlength{\baselineskip}{18pt}
\begin{titlepage}

\begin{flushright}
KOBE-TH-12-03\\%
HRI-P-12-01-001 
\end{flushright}
\vspace{1.0cm}
\begin{center}
{\Large\bf Anomalous Higgs Interactions\\[5pt] in Gauge-Higgs Unification} 
\end{center}
\vspace{25mm}

\begin{center}
{\large
K. Hasegawa, 
Nobuaki Kurahashi$^*$, 
C. S. Lim$^*$
and Kazuya Tanabe$^*$
}
\end{center}
\vspace{1cm}
\centerline{{\it
The Harish-Chandra Research Institute, Chhatnag Road, Jhusi, Allahabad 211 019, India.}}

\centerline{{\it
$^*$Department of Physics, Kobe University,
Kobe 657-8501, Japan.}}
%
%
\vspace{1.5cm}
\centerline{\large\bf Abstract}
\vspace{0.5cm}
We discuss anomalous Higgs interactions in the scenario of gauge-Higgs unification. 
In the scenario Higgs originates from higher dimensional gauge field and has a physical meaning as AB phase or Wilson loop.
As its inevitable consequence, physical observables are expected to be periodic in the Higgs field.
In particular, the Yukawa coupling is expected to show some periodic and non-linear behavior as the function of the Higgs VEV.
For a specific choice of the VEV, the Yukawa coupling of KK zero mode fermion even vanishes.
On the other hand, the Yukawa coupling is originally provided by gauge interaction, which is linear in the Higgs field.
We discuss how such two apparent contradiction about the non-linearity of the Yukawa coupling can be reconciled and at the same time how these two ``pictures" give different predictions in the simplest framework of the scenario: $SU(3)$ electroweak model in 5-dimensional flat space-time with orbifolding.
The deviation of the Yukawa coupling from the standard model prediction is also calculated for arbitrary VEV.
We study ``$H$-parity", which guarantees the stability of the Higgs for a specific choice of the VEV.
Also discussed is the Higgs interaction with $W^{\pm}$ and $Z^{0}$.
It turns out that in our framework of flat space-time the interaction does not show deviation from the standard model prediction, except for the specific case of the VEV.

\end{titlepage}

\newpage
\section{Introduction}

In spite of its great success especially in the sector of gauge interactions, the standard model still seems to have unsettled theoretical problems in its Higgs sector:
\begin{itemize}
\item The hierarchy problem  

It is well-known that the attempts to solve this problem, in particular the problem of quadratically ``divergent" quantum correction to the Higgs mass, have been main motives for various scenarios of physics beyond the standard model.

\item The origin of hierarchical fermion masses and flavor mixings 

Though fermion masses seem to show some regularity on their dependence on the generation number, the origin of the hierarchical fermion masses and flavor mixings have not been understood in a natural way.

\item The origin of CP violation  

In spite of the great success of Kobayashi-Maskawa model, the origin of CP violation still seems to be not conclusive yet.

\item The origin of Higgs itself  

\end{itemize}
These problems may stem  from the fact that there is no guiding principle (symmetry) to restrict the interactions of Higgs in the standard model.

In this paper we discuss gauge-Higgs unification (GHU) as a scenario of physics beyond the standard model.
In GHU, Higgs is identified with the Kaluza-Klein (KK) zero mode of extra space component of gauge field and thus the unification of 4-dimensional (4D) gauge and Higgs interactions is achieved in the framework of higher dimensional gauge theory.
The scenario itself is not new \cite{1979Manton, 1979Fairlie, 1983Hosotani}.
Importantly, the Hosotani mechanism for the dynamical gauge symmetry breaking due to the VEV of the extra-space component was proposed \cite{1983Hosotani}.

As the scenario of elementary particles, GHU, relying on higher dimensional gauge symmetry, is expected to shed some light on the problems listed above.
In fact, the quantum correction to the Higgs mass has been demonstrated to be finite by the virtue of higher dimensional gauge symmetry, once all KK modes are summed up in order to guarantee the extra dimensional gauge symmetry \cite{1998HIL}.  
Thus the GHU was realized to be viable as a model of elementary particles, since it provides a new avenue to solve the hierarchy problem without invoking SUSY and opens a new possibility of physics beyond the standard model.
In fact, the minimal $SU(3)$ unified electroweak model incorporating the standard model was constructed along this line of motivation \cite{KLY, SSS}.
Since then, much attention has been paid to the scenario and many interesting works have been done from various points of view \cite{2003BN, 2003CGM, GMN, 2004HHKY, HNT, 2005MSSS, 2004BQ, 2005PS, 2006CCP, PSW, MT, 2007LM, ACDP, 2005OW, HM, CPSW, 2007Hatanaka, 2007LM2, 2008Falkowski, 2008MO, 2009LM, 2009HSY, 2001ABQ, 2002GIQ, 2003CNP, 2008LMH, 2004HLM, HMTY, 2007ALM, 2000DHR, 2001g-2UED, 2009ALM, 2010ALM}.

Interestingly, GHU is closely related to other attractive scenarios aimed to solve the hierarchy problem, such as dimensional deconstruction \cite{ACG2001} and little Higgs model\cite{S2004}.
This is not surprising, since the theory of dimensional deconstruction can be regarded as 
a latticized GHU, where only extra space is latticized keeping extra dimensional gauge symmetry by use of link-variable.
It also should be emphasized that (the bosonic part of) point particle limit of open superstring theory, say 10-dimensional SUSY Yang-Mills theory is a sort of GHU.

In the context of the problems listed above, if the origin of Higgs is gauge boson, the following issues are challenging: 
\begin{itemize} 
\item to break CP 

\item to  realize fermion mass hierarchy  

\item to  accommodate flavor mixing 
\end{itemize} 
Let us note that in GHU the Yukawa coupling, being gauge coupling to start with, is real and universal among generations.
On the other hand, once these issues are settled, the scenario should provide us with new types of mechanisms for CP and flavor violation, and its predictions are expected to be predictive relying on the gauge principle.

{}From such point of view some works have been already completed on the subjects of flavor mixing and FCNC processes \cite{AKLMT}, which always has been a touchstone of various physics beyond the standard model, and CP violation \cite{2009ALM2, 2010LMN, 2012AKMT}.

While GHU relying on gauge principle may shed some lights on the long-standing problems of Higgs interactions, it is of crucial importance whether the scenario makes its characteristic predictions which are not shared by the standard model as the inevitable consequence of 
the fact that Higgs is a gauge boson.
{}From such point of view in this paper we discuss anomalous Higgs interaction in GHU.
Namely, we argue that in contrast to the case of the standard model, Yukawa coupling is non-diagonal, in general, even in the base of mass eigenstates of quarks and when focused on the zero KK mode sector, the Yukawa coupling deviates from that of the standard model and even vanishes in an extreme case.

Such anomalous Higgs interactions are known to be inevitable consequence of the Higgs as a gauge field.
To see this, let us begin with the fact that in gauge theories with spontaneous gauge symmetry breaking the fermion  mass term is generically written as 
\be 
\label{1.1}
m(v) \bar{\psi} \psi 
\ee
for a given mass eigenstate of fermion $\psi$, where $m(v)$ is a function of the VEV $v$ of Higgs field.
Physical Higgs field $h$ is a shift of the Higgs field from the VEV and therefore the interaction of $h$ with $\psi$ is naturally anticipated to be obtained by replacing $v$ by $v + h$.
This procedure works perfectly well for the standard model.
Namely, in the case of the standard model
\be 
\label{1.2} 
m(v) = fv\ ,
\ee
where $f$ is a Yukawa coupling constant, and the replacement $v \to v + h$ correctly gives the Yukawa interaction of $h$ with $\psi$:
\be 
\label{1.3}  
m(v+h)\bar{\psi}\psi = f(v + h) \bar{\psi}{\psi}\ .
\ee 
We also note the Yukawa coupling is given as the first derivative of the function : 
\be 
\label{1.4} 
f = \frac{d m(v)}{dv}\ .
\ee
So far everything seems to be just trivial.

We, however, realize that in GHU the situation is not trivial.
In GHU, our Higgs field is the zero-mode of some extra space component of gauge field $A_{y}^{(0)}$ (assuming 5D space-time).
Thus the VEV $v$ is a constant gauge field, which having vanishing field strength is usually regarded as unphysical, {\it i.e.} pure gauge.
However, in the case where the extra space is a circle $S^1$, a non-simply-connected space, the zero mode $A_y^{(0)}$ has a physical meaning as a Aharonov-Bohm (AB) phase or Wilson loop:
\be 
\label{1.5} 
W = P\,\e^{i\frac g2\!\oint\!A_ydy}
  = \e^{ig_4 \pi R A^{(0)}_{y}} ,
\ee   
where the integral is along $S^1$ and $g, g_4$ are 5D and 4D gauge couplings, respectively.
$R$ is the radius of $S^1$.
The integral $\oint\!A_y dy$ may be regarded as a magnetic flux $\Phi$ penetrating inside the circle,  
\be 
\label{1.6} 
g_{4}A_{y}^{(0)} =  g\frac{\Phi}{2\pi R},  
\ee
and therefore is physical and cannot be gauged away.

It is interesting to note that $W$ \eqref{1.5} is a periodic function of $A_y^{(0)}$.
Namely, in GHU, Higgs field appears in the form of \lq\lq non-linear realization".
Such periodicity in the Higgs field never appears in the standard model and therefore is expected to lead to quite characteristic prediction of GHU scenario.
Namely, as the characteristic feature of GHU we expect that physical observables have periodicity in the Higgs field:
\be 
\label{1.7}  
v ~~\lra~~ v + \frac{2}{g_{4}R}\ .
\ee
A similar thing happens in the quantization condition of magnetic flux in super-conductor:
\be 
\label{1.8}  
\Phi = \frac{2\pi}{e}n \qquad \big(\,n: \text{integer}\,\big)\ ,
\ee
where the unit of the quantization $\frac{2\pi}{e}$ corresponds to the period in \eqref{1.7}.
The effective potential as the function of the Higgs (VEV) is a typical example of the observables showing such periodicity:
\be 
\label{1.9}  
V(v) \propto \frac{3}{4\pi^{2}}\frac{1}{(2\pi R)^{4}} 
\sum_{n= 1}^{\infty} \frac{\cos (n g_{4}\pi R v)}{n^{5}}, 
\ee 
which is the simplified formula for the contributions of the fields with vanishing bulk masses.

We expect that the mass eigenvalue in \eqref{1.1} also has the periodicity.
In fact in this paper we will show that the mass eigenvalues for light zero-mode quarks with ``$Z_2$-odd" bulk masses are well approximated by
\be 
\label{1.10}  
m(v) \propto \sin\!\left(\frac{g_4}{2}\pi R v\right),
\ee 
which leads to a Higgs interactions with quarks, behaving as trigonometric function of $h$ and therefore non-linear interactions\,!
Namely,
\be 
\label{1.11}  
m(v + h) \propto \sin\!\left\{\frac{g_{4}}{2}\pi R (v+h)\right\}
\ee
and the Yukawa coupling, {\it i.e.} the coupling of the linear interaction of Higgs $h\bar\psi\psi$, is given as
\be  
\label{1.12}
f = \frac{dm(v)}{dv} \propto \cos\!\left(\frac{g_{4}}{2}\pi R v\right).
\ee 
We now realize that the Yukawa coupling even vanishes for an extreme case of
\be  
\label{1.13}
x \equiv \frac{g_{4}}{2}\pi R v = \frac{\pi}{2}\ .
\ee

This kind of ``anomalous" Higgs interaction has been first pointed out in curved Randall-Sundrum (RS) 5D space-time and for the gauge group $SO(5) \times U(1)$ \cite{2007HS, 2008HOOS, 2009HK, 2009HKT, 2010HTU}.
Even the possibility that the Higgs, being rather stable, plays the role of dark matter has 
been pointed out \cite{2009HKT}.

\begin{figure}[!t]
\centering
\begin{tabular}{ccc}
\includegraphics[bb= 0 0 258 163, scale=0.6]{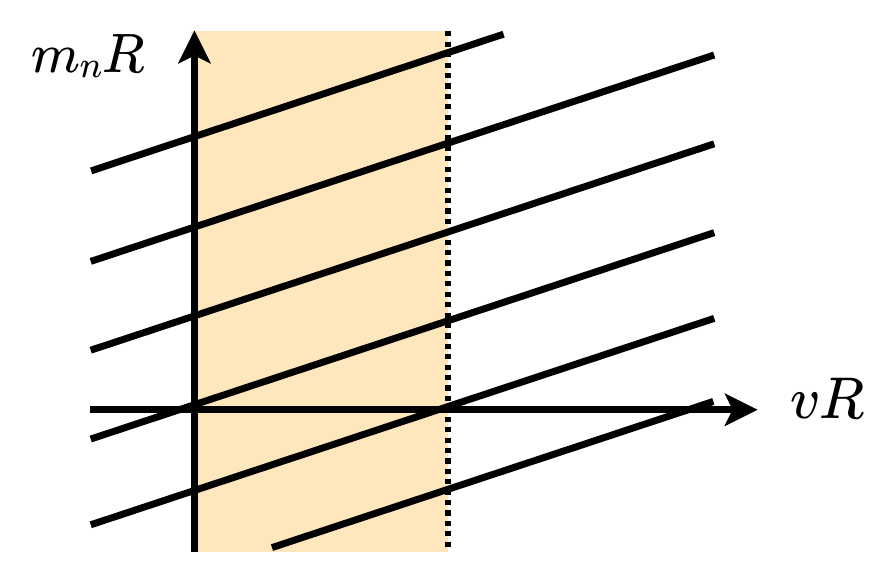}
	\qquad & \qquad
\includegraphics[bb= 0 0 262 164, scale=0.6]{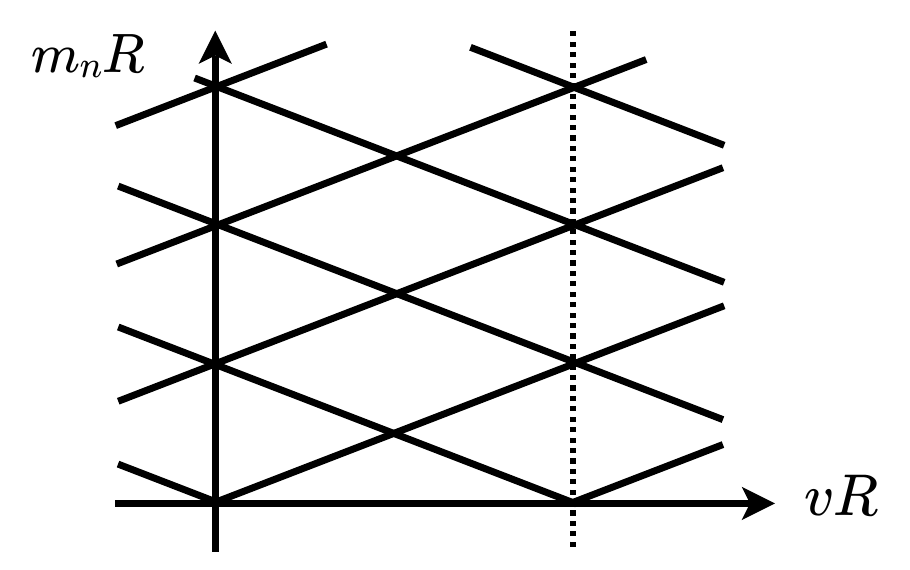}\\
\text{(a)} \qquad&\qquad
\text{(b)}
\end{tabular}
\caption{\small%
(a) : KK mass eigenvalues of fermion.
(b) : The eigenvalues after chiral transformation.
}
\label{fig:levelcross}
\end{figure}

\begin{figure}[!t]
\centering
\begin{tabular}{ccc}
\includegraphics[bb= 0 0 248 183, scale=0.6]{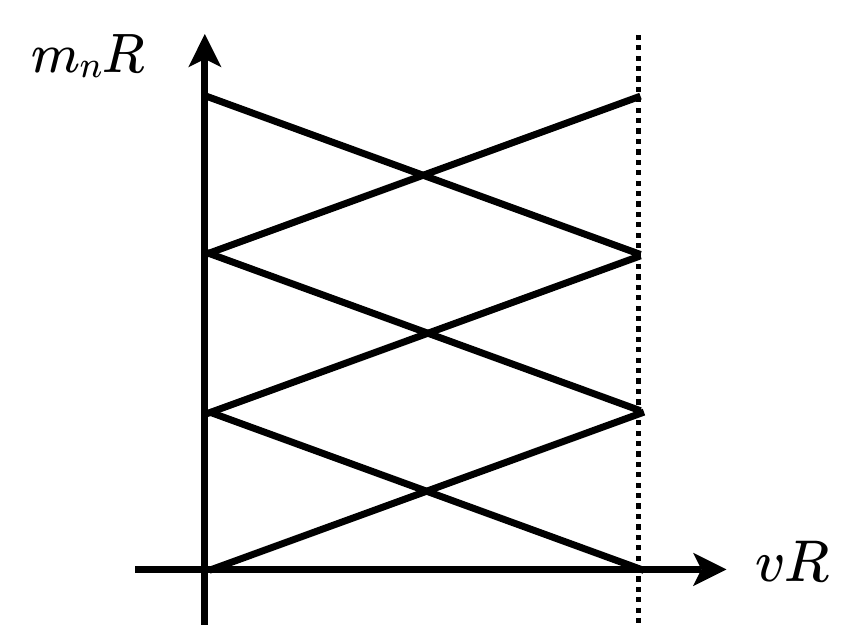}
	\qquad & \qquad
\includegraphics[bb= 0 0 249 183, scale=0.6]{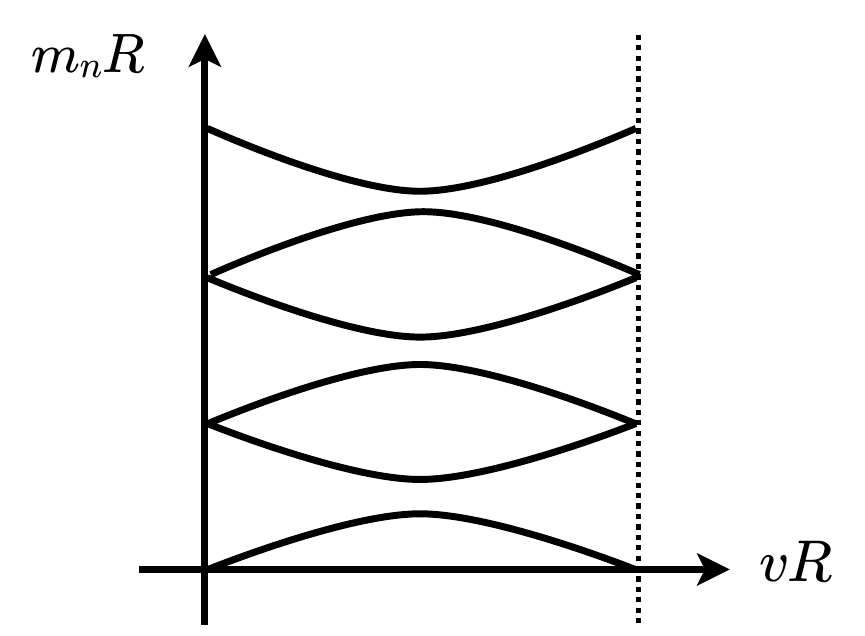}\\
\text{(a)} \qquad&\qquad
\text{(b)}
\end{tabular}
\caption{\small%
(a) : \lq\lq Level crossing" among mass eigenvalues ($M = 0$).
(b) : The level crossing is avoided by the shift of degenerate mass eigenvalues of $\mathcal O(M)$ ($M \ne 0$).
}
\label{fig:levelcross2}
\end{figure}

We, however, know that the Yukawa interaction given in the original lagrangian does not have such non-linearity and is linear in the physical Higgs field $h$, just as in the standard model:
\be 
	\label{1.14} 
	\bar{\psi}
	\left\{
	i\partial_\mu\gamma^\mu-\gamma_5\partial_y
	+i\gamma_5g_4\frac{\lambda_6}2(v + h)-M\epsilon(y)
	\right\}
	\psi\ ,
\ee
which is the relevant part in the $SU(3)$ model we discuss later and $\lambda_{6}$ is a Gell-Mann matrix.
In fact, the KK mass eigenvalues for a specific case of vanishing bulk mass $M$ are known to be linear in $v$:
\be 
\label{1.15} 
m_n
  = \frac nR + \frac{g_4}2v \qquad \big(\,n: \text{integers}\,\big).
\ee 
In this specific case, although the eigenvalues themselves are linear in $v$, the mass spectrum as the whole is known to be periodic as is seen in \autoref{fig:levelcross} (a).
We note that in this case the Yukawa coupling given by \eqref{1.4} is just a constant as in the standard model, except the specific situation $x = \frac\pi2$.
In \autoref{fig:levelcross} (b), which is obtained from (a) by chiral transformations for negative KK modes $n<0$ \big(see \eqref{mnforM0case}\big), there appears a level crossing at $x = \frac\pi2$ and the derivative cannot be defined.
Though we expect that the level crossing is lifted once the mixing among the crossing two KK modes is taken into account, the mixing seems not to be allowed for vanishing bulk mass, because of the conservation of extra space component of momentum.
We will see later that by introducing the bulk mass $M$ the level crossing is avoided as is shown in \autoref{fig:levelcross2} (b).
This may be understood as the result of the violation of translational invariance in the extra space due to the introduction of the bulk mass.

At the first glance, these two viewpoints or ``pictures", {\it i.e.} the one which claims non-linear Higgs interactions as is shown in \eqref{1.11} and the other one which claims linear Yukawa interaction of $h$ as is shown in \eqref{1.14}, seem to be contradictory with each another.
Both pictures, however, are based on some reliable arguments and there should be a way to reconcile these two.

Hence, the main purpose of this paper is to study the interesting properties of anomalous interactions, in particular to clearly understand how these two pictures are reconciled with each another, in the simplest framework of GHU, {\it i.e.}~$SU(3)$ electroweak gauge model in 5D space-time with an orbifold $S^1\!/Z_2$ as its extra space \cite{KLY, SSS}.
As the matter field we introduce a $SU(3)$ triplet fermion.
We are also interested in the issue whether these two pictures make different predictions in some range of supposed energies.

It will be shown that the Higgs interaction with fermion is linear in $h$ as is seen in \eqref{1.14} and can be written in the form of matrix in the base of fermion's 4D mass eigenstates, {\it i.e.} KK modes.
In contrast to the case of the standard model, the ``Yukawa coupling matrix" is generally non-diagonal.
For instance in the specific case $x = \frac\pi2$, all diagonal elements are known to disappear and the matrix becomes completely off-diagonal.
The mass function $m(v+h)$ such as \eqref{1.11} is nothing but the eigenvalue of the 4D mass operator for the zero-mode fermion,  where $h$ is regarded as a constant on an equal footing with the VEV $v$.
Namely, it is an eigenvalue of the matrix in the base of all KK modes, obtained from the $y$-integral ($y$ is an extra space coordinate) of the free lagrangian \eqref{1.14} with the 4D
kinetic term being ignored:
\be 
	\label{1.16} 
	\int_{-\pi R}^{\pi R}\hspace{-5mm}dy\,
	\bar{\psi}
	\left\{\gamma_5\partial_y-i\gamma_5g_4\frac{\lambda_6}2(v+h)+M\epsilon(y)\right\}
	\psi\ .
\ee 
As long as the Yukawa coupling matrix, which is the part linear in $h$ in \eqref{1.16} has off-diagonal elements, the eigenvalues of the matrix obtained from \eqref{1.16} can be non-linear in $h$. 
Thus the two pictures are not contradictory with each another.
On the other hand, we will point out that the predictions for the quadratic Higgs interactions in two pictures show some difference when Higgs mass and/or Higgs 4-momentum 
cannot be ignored, which reasonably may be the case in the situation of LHC experiment or future linear collider.

In addition, the ``$H$-parity" proposed in \cite{2009HKT, 2010HTU} to implement the stability of the Higgs at $x = \frac\pi2$ is investigated from our own viewpoint in our model.
Also discussed is the Higgs interaction with massive zero mode gauge bosons $W^\pm$ and $Z^0$.

In section 2, our model is briefly described and quark mass eigenvalues and corresponding mode functions are derived.
In section 3, anomalous Higgs interaction with quarks is discussed.
First by use of the wisdom of quantum mechanics we argue that two pictures can be 
reconciled with each another.
By use of such wisdom we point out that the Yukawa coupling of the Higgs with the zero mode $d$ quark can be calculated in two different ways and we confirm by explicit calculations that these two methods provide exactly the same result.
At the same time we point out that two pictures make different predictions on the quadratic Higgs interaction with the quark under some circumstance.
The formula to give the deviation of the anomalous Yukawa coupling from the standard model prediction for an arbitrary Higgs VEV is obtained and an approximated formula for light quarks is shown to be in good agreement with the exact result.
In section 4, $H$-parity is discussed and we show that only in the specific case of $x = \frac{\pi}{2}$ the parity symmetry is not broken spontaneously, and therefore meaningful.
In section 5, we address the issue of Higgs interaction with massive gauge bosons $W^{\pm}$ and $Z^{0}$.
We show that except for the specific case $x = \frac{\pi}{2}$ the Higgs interaction is always linear and there is no deviation from the standard model prediction, in contrast to the result in \cite{2007HS, 2008HOOS}.

\section{The Model}
The model we take is $SU(3)$ GHU model
with triplet fermion as the matter field
\begin{align}
	\Psi
  = \left[\begin{array}{c}
	 \psi_1\\[2pt]
	 \psi_2\\[2pt]
	 \psi_3
	\end{array}\right]
\end{align}
with the orbifolding condition
\begin{align} 
\label{2.1} 
	\Psi ~~\lra~~ \gamma^5P\Psi
	\qquad , \qquad
	P
  = \left[\begin{array}{ccc}
	 1 & 0 & 0\\[2pt]
	 0 & 1 & 0\\[2pt]
	 0 & 0 & -1\!\!
	\end{array}\right]\ .
\end{align}
The zero mode may be regarded as Weyl fermions of quarks: when the VEV $v$ can be ignored,  
\begin{align} 
\label{fermionzeromode}
	\Psi^{(0)}
  = \left[\begin{array}{c}
	 u_L\\[2pt]
	 d_L\\[2pt]
	 d_R
	\end{array}\right]\ .
\end{align}

What we are interested in is the 4D mass term and Yukawa interaction of $\Psi$,
whose relevant lagrangian is given as
\begin{align}
	\label{5dlag}
	\mathcal L
  = \bar\Psi
	\left\{
	i\partial_\mu\gamma^\mu
	+\varGamma^5\!
	 \left(
	 i\partial_y
	 +\frac{g_5}2A_y^{6(0)}\!
	  \left[\begin{array}{ccc}
	   0 & 0 & 0\\[2pt]
	   0 & 0 & 1\\[2pt]
	   0 & 1 & 0
	  \end{array}\right]
	 \right)\!
	-M\epsilon(y)
	\right\}
	\Psi
	~\quad\Big(\,\varGamma^5 = i\gamma^5\,\Big)\ ,
\end{align}
where $\epsilon(y)$ is the sign function
\begin{align}
	\epsilon(y)
  = \left\{
	\begin{array}{ccc}
	 +1 & \text{for }\,y>0\\[4pt]
	 -1 & \text{for }\,y<0
	\end{array}\right.
\end{align}
and where $A_y^{6(0)}$ denotes the zero mode of $A_y^6$
and is identified with the neutral component of Higgs doublet:
\begin{align}
	\label{5dvev}
	A_y^{6(0)} = v_5+H\ .
\end{align}
In \eqref{5dvev}, $g_5v_5 = g_4v$ \big($g_5$, $g_4$ : 5D \& 4D gauge couplings\big)
with $v$ being 4D VEV of Higgs and $h$ given by	$H = \frac1{\sqrt{2\pi R}}h$ is nothing but our Higgs field.

Since in this model only $d$-quark gets its mass and Yukawa interaction with Higgs $h$,
we focus on the subspace of $\Psi$,
\begin{align}
	\psi
  = \left[\begin{array}{c}
	 \psi_2\\[2pt]
	 \psi_3
	\end{array}\right]\ ,
\end{align}
whose free lagrangian is read off from \eqref{5dlag} as
\begin{align}
	\label{freelag}
	\mathcal L_{\rm free}
  = \bar\psi
	\Big\{
	i\partial_\mu\gamma^\mu
	-\gamma^5\!\left(\partial_y-i\frac{g_4}2v\sigma_1\right)-M\epsilon(y)
	\Big\}
	\psi\ ,
\end{align}
where $\sigma_1$ is one of Pauli matrices.
The orbifolding condition is imposed on $\psi_2$ and $\psi_3$ as
\begin{subequations}
\begin{align}
	\psi_{2L}(x,-y)
 &= +\psi_{2L}(x,y)\qquad , \qquad
	\psi_{2R}(x,-y)
  = -\psi_{2R}(x,y)\ ,\\
	\psi_{3L}(x,-y)
 &= -\psi_{3L}(x,y)\qquad , \qquad
	\psi_{3R}(x,-y)
  = +\psi_{3R}(x,y)\ .
\end{align}
\end{subequations}
The Weyl spinors $\psi_{2L}$, $\psi_{2R}$ and $\psi_{3L}$, $\psi_{3R}$ are regarded
as periodic continuous functions of $y$:
\begin{align}
	\label{continbc}
	\psi_{2L}(x,-\pi R) = \psi_{2L}(x,\pi R)
	\qquad , \qquad
	\psi_{2R}(x,-\pi R) = \psi_{2R}(x,\pi R)
	\quad , \quad \text{etc.}
\end{align}


\subsection{The equations of motion and mode functions for fermion}

The equation of motion for $\psi$ obtained from \eqref{freelag} is
\begin{align}
	\label{eompsi}
	\Big\{
	i\partial_\mu\gamma^\mu
	-\gamma^5\!\left(\partial_y-i\frac{g_4}2v\sigma_1\right)-M\epsilon(y)\Big\}\psi
  = 0\ .
\end{align}
Because of the presence of $\sigma_1$ term, the equation becomes coupled equation.
To remedy this, we define $\hat\psi$ so that
\begin{align}
	\label{hatpsi}
	\psi = \exp\!\Big\{i\frac{g_4}2vy\sigma_1\Big\}\hat\psi\ .
\end{align}
Then in terms of $\hat\psi$, the $\sigma_1$ term disappears in its equation of motion:
\begin{align}
	\label{eom}
	\Big\{i\partial_\mu\gamma^\mu-\gamma^5\partial_y-M\epsilon(y)\Big\}\hat\psi = 0\ .
\end{align}

Though the effect of VEV disappears in \eqref{eom},
on the other hand the boundary condition of $\hat\psi$ is no longer periodic.
Namely, from \eqref{continbc} and \eqref{hatpsi},
\begin{align}
	\label{twistpbc}
	\hat\psi(x,-\pi R) = \langle W \rangle \hat\psi(x,\pi R)
	\qquad\text{where}\qquad
	\langle W \rangle \equiv \e^{i\pi Rg_4v\sigma_1}\ .
\end{align}
$\langle W\rangle$ is nothing but the VEV of the ``Wilson loop", or AB phase.
Writing
\begin{align}
	\label{subhatpsi}
	\hat\psi
  = \left[\begin{array}{c}
	 \hat\psi_2\\[2pt]
	 \hat\psi_3
	\end{array}\right]\ ,
\end{align}
it is interesting to note that $\hat\psi_2$, $\hat\psi_3$ obey the same orbifolding condition as $\psi_2$, $\psi_3$:
\begin{subequations}
\begin{align}
	\label{ochatpsi2}
	\hat\psi_{2L}(x,-y)
 &= +\hat\psi_{2L}(x,y)\qquad , \qquad
	\hat\psi_{2R}(x,-y)
  = -\hat\psi_{2R}(x,y)\ ,\\
	\label{ochatpsi3}
	\hat\psi_{3L}(x,-y)
 &= -\hat\psi_{3L}(x,y)\qquad , \qquad
	\hat\psi_{3R}(x,-y)
  = +\hat\psi_{3R}(x,y)\ .
\end{align}
\end{subequations}
This is because in the relation \eqref{hatpsi}
\begin{align}
	\e^{i\frac{g_4}2vy\sigma_1}
  = \cos\!\left(\frac{g_4}2vy\right)\!\bs1_{2\times2}
	+i\sin\!\left(\frac{g_4}2vy\right)\!\sigma_1\ .
\end{align}

Eq.\,\eqref{eom} shows that $\hat\psi_2$ and $\hat\psi_3$ obey the same equation of motion.
We first focus on the equation for $\hat\psi_2$:
\begin{align}
	\label{eom2}
	\Big\{i\partial_\mu\gamma^\mu-\gamma^5\partial_y-M\epsilon(y)\Big\}\hat\psi_2 = 0\ .
\end{align}
Let us expand $\hat\psi_2$ in terms of mode functions, which are eigenfunctions with definite 4D mass eigenvalues $m_n$:
\begin{align}
	\label{mepsi2}
	\hat\psi_2(x,y)
  = \sum^\infty_{n=0}f_{Le}^{(n)}(y)\hat\psi_{2L}^{(n)}(x)
	+\sum^\infty_{n=0}f_o^{(n)}(y)\hat\psi_{2R}^{(n)}(x)\ .
\end{align}
The mode functions $f_{Le}^{(n)}$ and $f_o^{(n)}$ are even and odd functions of $y$:
\begin{align}
	 f_{Le}^{(n)}(-y) &= +f_{Le}^{(n)}(y)
	 \quad , \qquad
	 f_o^{(n)}(-y) = -f_o^{(n)}(y)
\end{align}
to be consistent with \eqref{ochatpsi2}.
$\hat\psi_{2L}^{(n)}$, $\hat\psi_{2R}^{(n)}$ are 4D Weyl spinors.
Applying the left-handed projection $L$ from the left of \eqref{eom2} and using \eqref{mepsi2}, we get
\begin{align}
	\label{kkmfrela}
	\Big\{i\partial_\mu\gamma^\mu\hat\psi_{2R}^{(n)}(x)\Big\}f_o^{(n)}(y)
	-\hat\psi_{2L}^{(n)}(x)\Big\{\partial_y+M\epsilon(y)\Big\}f_{Le}^{(n)}(y)
  = 0\ .
\end{align}
On the other hand,
Dirac equation for $\hat\psi_2^{(n)}(x) = \hat\psi_{2R}^{(n)}(x)+\hat\psi_{2L}^{(n)}(x)$ is written as
\begin{align}
	\label{kkmfrelaforde}
	(i\sls\partial-m_n)\hat\psi_2^{(n)}(x) = 0
	\qquad \lra \qquad
	i\sls\partial\hat\psi_{2R}^{(n)}(x)-m_n\hat\psi_{2L}^{(n)}(x)
  = 0\ .
\end{align}
We thus conclude, by comparing \eqref{kkmfrela} and \eqref{kkmfrelaforde},
\begin{subequations} \label{susyrelaofmf}
\begin{align}
	\label{susyrelaofmf1}
	\Big\{\partial_y+M\epsilon(y)\Big\}f_{Le}^{(n)}(y)
  = m_nf_o^{(n)}(y)\ .
\end{align}
Similar argument yields
\begin{align}
	\label{susyrelaofmf2}
	\Big\{\!-\!\partial_y+M\epsilon(y)\Big\}f_o^{(n)}(y)
  = m_nf_{Le}^{(n)}(y)\ .
\end{align}
\end{subequations}
Equations of \eqref{susyrelaofmf} imply the presence of Quantum Mechanical SUSY \cite{2005LNSS}.
Namely ``super-charge" $Q$ and ``Hamiltonian" $H$ may be defined as follows in the base of \big($f_{Le}^{(n)}$, $f_o^{(n)}$\big);
\begin{subequations}
\begin{align}
	Q
  = \left[\begin{array}{ccc}
	 0 & -\partial_y+M\epsilon(y)\\[3pt]
	 \partial_y+M\epsilon(y) & 0
	\end{array}\right], 
\end{align}
\begin{align}
	H
 &= Q^2\notag\\
 &= \left[\begin{array}{ccc}
	 -\partial_y^2+M^2-2M\Big\{\delta(y)-\delta(y-\pi R)\Big\} & 0\\[5pt]
	 0 & -\partial_y^2+M^2+2M\Big\{\delta(y)-\delta(y-\pi R)\Big\}
	\end{array}\right].
\end{align}
\end{subequations}
Hence the commutator of $Q$ and $H$ vanishes indicating the presence of the supersymmetry:
\begin{align}
	\Big[\, Q\,,\,H\,\Big]
  = 0\ .
\end{align}

Actually, the ``Hamiltonian" $H$ just corresponds to the 4D mass-squared operator for fermions.
Namely, eqs. \eqref{susyrelaofmf1} and \eqref{susyrelaofmf2} are combined as,
\begin{subequations} \label{msoperator}
\begin{align}
	\Big\{-\!\partial_y+M\epsilon(y)\Big\}
	\Big\{\partial_y+M\epsilon(y)\Big\}
	f_{Le}^{(n)}(y)
 &= m_n^2f_{Le}^{(n)}(y)\notag\\
	\label{msoperator1}
	\lra~~
	\Big[-\!\partial_y^2+M^2-2M\Big\{\delta(y)-\delta(y-\pi R)\Big\}\Big]
	f_{Le}^{(n)}(y)
 &= m_n^2f_{Le}^{(n)}(y)\ .
\end{align}
Similarly, for $f_o^{(n)}(y)$,
\begin{align}
	\label{msoperator2}
	\Big[-\!\partial_y^2+M^2+2M\Big\{\delta(y)-\delta(y-\pi R)\Big\}\Big]
	f_o^{(n)}(y)
 &= m_n^2f_o^{(n)}(y)\ .
\end{align}
\end{subequations}

The mode functions $f_{Le}^{(n)}$ and $f_o^{(n)}$ and the eigenvalue $m_n$ are determined so that they satisfy the following relations:
\begin{enumerate}
\renewcommand{\theenumi}{(\roman{enumi})}
\item
Equation of motion in the bulk
\begin{subequations}
	\label{psi2condition}
\begin{align}
	\label{eombulk}
\left\{\begin{array}{rl}
 \left(-\partial_y^2+M^2\right)\!f_{Le}^{(n)}(y) &\!\!\!\!= m_n^2f_{Le}^{(n)}(y)\\[5pt]
 \left(-\partial_y^2+M^2\right)\!f_o^{(n)}(y) &\!\!\!\!= m_n^2f_o^{(n)}(y)
\end{array}\right.
\qquad\Big(\,\text{for }\,0<|y|<\pi R\,\Big)\ .
\end{align}

\item
Continuity at the fixed point $y=0$
\begin{align}
	\label{y0continuity}
	f_{Le}^{(n)}, f_o^{(n)} \text{ are continuous for } 0\leq |y| < \pi R\ .	
\end{align}

\item
Discontinuity of the derivative at the fixed point $y=0$\,%
\footnote{%
The discontinuity condition (iii) for $f_o^{(n)}$ is trivial,
since its derivative is even function without discontinuity and $f_o^{(n)}(0) = 0$
}

By integrating \eqref{msoperator1} and \eqref{msoperator2}
in the infinitesimal regions $-\varepsilon \leq y \leq \varepsilon$,
we get
\begin{align}
	\label{disconofderiv}
	\lim_{\varepsilon\to+0}\left(\partial_yf_{Le}^{(n)}\right)\!(y=\varepsilon)
	-\lim_{\varepsilon\to-0}\left(\partial_yf_{Le}^{(n)}\right)\!(y=\varepsilon)
  = -2Mf_{Le}^{(n)}(0)\ .
\end{align}
\end{subequations}

\end{enumerate}
There are two more conditions to be imposed, {\it i.e.} the continuity of $\psi$ at another fixed point $y=\pm\pi R$, {\it i.e.}~\eqref{twistpbc}, and the condition similar to \eqref{disconofderiv} at $y=\pm\pi R$.
We, however, realize from \eqref{twistpbc} and \eqref{subhatpsi}, that, except for $v=0$ \big($W=\bs1_{2\times2}$\big), $\hat\psi$ is discontinuous.
Thus the condition similar to \eqref{disconofderiv} should be imposed on $\psi$ itself, not $\hat\psi$.
We also note that except for $v=0$, $\frac1{g_4R}$ the condition \eqref{twistpbc} causes the mixing between $\hat\psi_2$ and $\hat\psi_3$.

Thus we now discuss the mode-expansion of $\hat\psi_3$.
{}From the $Z_2$-parity assignment \eqref{ochatpsi3},
the mode-expansion is as follows:
\begin{align}
	\hat\psi_3(x,y)
  = \sum^\infty_{n=0}f_o^{(n)}(y)\hat\psi_{3L}^{(n)}(x)
	+\sum^\infty_{n=0}f_{Re}^{(n)}(y)\hat\psi_{3R}^{(n)}(x)\ .
\end{align}
Similarly to \eqref{susyrelaofmf}, the relations between two kinds of mode functions are
\begin{subequations}
	\label{susyrelaofmfb}
\begin{align}
	\label{susyrelaofmf12}
	\Big\{\partial_y+M\epsilon(y)\Big\}f_o^{(n)}(y)
 &= m_nf_{Re}^{(n)}(y)\ ,\\  
	\label{susyrelaofmf22}
	\Big\{\!-\!\partial_y+M\epsilon(y)\Big\}f_{Re}^{(n)}(y)
 &= m_nf_o^{(n)}(y)\ .
\end{align}
\end{subequations}
Comparing with \eqref{susyrelaofmf}, we readily know that $f_{Re}^{(n)}(y)$ and $f_o^{(n)}(y)$ are easily obtained by the replacement
\begin{align} 
\label{replacement} 
	f_{Le}^{(n)}(y) ~\lra~ - f_{Re}^{(n)}(y)
	\quad , \qquad
	M ~\lra~ -\!M\ .
\end{align}

The mode function $f_o^{(n)}(y)$, $f_{Le}^{(n)}(y)$ and $f_{Re}^{(n)}(y)$ are easily derived
by the following procedure.
Let us note that the discontinuity at $y = 0$ is not applicable for the odd function $f_o^{(n)}(y)$.
Therefore, $f_o^{(n)}(y)$ should be just a continuous function for $|y|<\pi R$,
satisfying \eqref{eombulk}.
Thus the mode function is easily obtained as
\begin{align} 
	\label{omodefunction}
	f_o^{(n)}(y)
  = \frac1{\sqrt{\pi R}}\sin\!\left(\sqrt{m_n^2-M^2}y\right)
	\qquad\big(\,n\geq0\,\big)\ , 
\end{align}
where the normalization factor is for $v = 0$ and the factor is corrected later for general value of $v$. 
To obtain even mode functions in a similar way is a little tedious because of \eqref{disconofderiv}.
Instead, we can utilize the relation of Quantum Mechanical SUSY.
Namely, by use of \eqref{susyrelaofmf2} and \eqref{omodefunction}, $f_{Le}^{(n)}(y)$ is easily derived (for $m_n>0$) as
\begin{align}
	\label{lemodefunction}
	f_{Le}^{(n)}(y)
 &= \frac1{m_n}\big\{\!-\!\partial_y+M\epsilon(y)\big\}f_o^{(n)}(y)
  = -\frac1{\sqrt{\pi R}}\cos\!\left(\sqrt{m_n^2-M^2}|y|+\alpha_n\right)\ ,
\end{align}
where $\alpha_n$ is defined as
\begin{align} \label{alphan}
	\cos\alpha_n = \frac{\sqrt{m_n^2-M^2}}{m_n}
	\qquad , \qquad
	\sin\alpha_n = \frac M{m_n}\ .
\end{align}
$f_{Re}^{(n)}(y)$ is given by utilizing the replacement \eqref{replacement}:
\begin{align} \label{remodefunction}
	f_{Re}^{(n)}(y)
  = \frac1{\sqrt{\pi R}}\cos\!\left(\sqrt{m_n^2-M^2}|y|-\alpha_n\right)\ .
\end{align}
It is interesting to note that $f_{Le}^{(n)}$ and $f_{Re}^{(n)}$ automatically satisfy \eqref{disconofderiv} and its counterpart.

The mode functions for the zero-modes, $f_{Le}^{(0)}$, $f_{Re}^{(0)}$ cannot be obtained from $f_o^{(n)}$ for the specific case of $v=0$, since in this case $m_0 = 0$ and the SUSY transformation cannot be used.
Instead, we directly solve \eqref{susyrelaofmf1} and \eqref{susyrelaofmf22} for $m_n = 0$ to get:
\begin{subequations} \label{0modefunction}
\begin{align}
	f_{Le}^{(0)}(y)
 &= \sqrt{\frac M{1-\e^{-2\pi RM}}}\e^{-M|y|}
	\quad , \qquad
	f_{Re}^{(0)}(y)
  = \sqrt{\frac M{1-\e^{-2\pi RM}}}\e^{-M(\pi R-|y|)}\ .
\end{align}
\end{subequations}

\subsection{The boundary condition at $\bs{|y| = \pi R}$ and mass spectra for fermion}

It has been shown that the boundary conditions at $y=0$, \eqref{y0continuity} and \eqref{disconofderiv}, are satisfied by the mode functions \eqref{omodefunction} and \eqref{lemodefunction}.
Strictly speaking, these conditions should be satisfied by $\psi$, not by $\hat\psi$.
We, however, realize that at $y=0$ the conditions for $\psi$ are identical with those for $\hat\psi$, since $\e^{i\frac{g_4}2vy\sigma_1} = \bs1_{2\times2}$ for $y=0$ and $\e^{i\frac{g_4}2vy\sigma_1}$ is a continuous function.

The situation changes for the case of another fixed point $|y| = \pi R$, since now the factor $\e^{i\frac{g_4}2vy\sigma_1}$ is no longer an identity.
The boundary conditions are now given as
\begin{itemize}
\item
Continuity of $\psi$ at $|y|=\pi R$
\begin{align}
	\label{continuitypiR}
 &  \psi(y=\pi R)-\psi(y=-\pi R) = 0\notag\\
	\lra~~
 &  \e^{i\frac{g_4}2v\pi R\sigma_1}\hat\psi(y=\pi R)
	-\e^{-i\frac{g_4}2v\pi R\sigma_1}\hat\psi(y=-\pi R) = 0\ .
\end{align}
As was mentioned in \eqref{twistpbc}, this relation can be written in terms of Wilson loop $W$.
Let us also note that \eqref{continuitypiR} is equivalent to demanding the equation,
\begin{align}
	\label{continuitypiR2}
	\e^{i\frac{g_4}2vy\sigma_1}\hat\psi\big|_{\rm odd} = 0
	\qquad \text{for } y = \pi R\ ,
\end{align}
where $\e^{i\frac{g_4}2vy\sigma_1}\hat\psi\big|_{\rm odd}$ denotes the odd function part of $\e^{i\frac{g_4}2vy\sigma_1}\hat\psi$.

\item
Discontinuity of $\partial_y\psi$ at $|y|=\pi R$
\begin{align}
	\label{discontinuitypiR}
 &  \gamma^5
	\Big\{
	\big(\partial_y\psi\big)(y=\pi R)-\big(\partial_y\psi\big)(y=-\pi R)
	\Big\}
  = -M\Big\{\epsilon(\pi R)-\epsilon(-\pi R)\Big\}\psi(\pi R)\notag\\
	\lra~
 &  \gamma^5
	\Big\{
	\e^{i\frac{g_4}2v\pi R\sigma_1}\big(\partial_y\hat\psi\big)(y=\pi R)
	-\e^{-i\frac{g_4}2v\pi R\sigma_1}\big(\partial_y\hat\psi\big)(y=-\pi R)
	\Big\}\notag\\
 &= -2M\e^{i\frac{g_4}2v\pi R\sigma_1}\hat\psi(y=\pi R)\ , 
\end{align}
which is equivalent to demanding that
\begin{align}
	\label{discontinuitypiR2}
	\gamma^5\e^{i\frac{g_4}2v\pi R\sigma_1}
	\big(\partial_y\hat\psi\big)\big|_{\rm odd}(y=\pi R)
  = -M\e^{i\frac{g_4}2v\pi R\sigma_1}\hat\psi(y=\pi R)\ .
\end{align}

\end{itemize}

Let us derive the conditions in order to determine the mass eigenvalue $m_n$ and corresponding mass eigenstate.
For a fixed KK mode $n$ with eigenvalue $m_n$, the eigenstate $\psi$ is written as
\begin{align}
	\psi
 &= \e^{i\frac{g_4}2vy\sigma_1}\hat\psi
  = \Big\{
	\cos\!\left(\frac{g_4}2vy\right)\!\bs1_{2\times2}
	+i\sin\!\left(\frac{g_4}2vy\right)\!\sigma_1\Big\}\!\!
	\left[\begin{array}{c}
	 f_{Le}^{(n)}(y)\hat\psi_{2L}^{(n)}(x)+f_o^{(n)}(y)\hat\psi_{2R}^{(n)}(x)\\[6pt]
	 f_o^{(n)}(y)\hat\psi_{3L}^{(n)}(x)+f_{Re}^{(n)}(y)\hat\psi_{3R}^{(n)}(x)
	\end{array}\right].
\end{align}
Thus, the condition \eqref{continuitypiR2} reads as
\begin{itemize}
\item
left-handed part
\begin{align}
	\label{piRlhs}
 &  \cos\!\left(\frac{g_4}2v\pi R\right)\!f_o^{(n)}(\pi R)\hat\psi_{3L}^{(n)}(x)
	+i\sin\!\left(\frac{g_4}2v\pi R\right)\!f_{Le}^{(n)}(\pi R)\hat\psi_{2L}^{(n)}(x)
  = 0\notag\\
	\lra~~
 &  \cos\!\left(\frac{g_4}2v\pi R\right)\!\sin\varphi_n\cdot\hat\psi_{3L}^{(n)}(x)
	-i\sin\!\left(\frac{g_4}2v\pi R\right)\!
	 \cos(\varphi_n+\alpha_n)\cdot\hat\psi_{2L}^{(n)}(x)
  = 0\ ,
\end{align}
where
\begin{align}
	\varphi_n \equiv \sqrt{m_n^2-M^2}\pi R\ .
\end{align}

\item
right-handed part
\begin{align}
	\label{piRlhsr}
 &  \cos\!\left(\frac{g_4}2v\pi R\right)\!f_o^{(n)}(\pi R)\hat\psi_{2R}^{(n)}(x)
	+i\sin\!\left(\frac{g_4}2v\pi R\right)\!f_{Re}^{(n)}(\pi R)\hat\psi_{3R}^{(n)}(x)
  = 0\notag\\
	\lra~~
 &  \cos\!\left(\frac{g_4}2v\pi R\right)\!\sin\varphi_n\cdot\hat\psi_{2R}^{(n)}(x)
	+i\sin\!\left(\frac{g_4}2v\pi R\right)\!
	 \cos(\varphi_n-\alpha_n)\cdot\hat\psi_{3R}^{(n)}(x)
  = 0\ .
\end{align}

\end{itemize}
Similarly the condition \eqref{discontinuitypiR2} reads as
\begin{itemize}
\item
left-handed part
\begin{align}
	\label{piRlhs2}
    \cos\!\left(\frac{g_4}2v\pi R\right)\!\sin\varphi_n\cdot\hat\psi_{2L}^{(n)}(x)
	+i\sin\!\left(\frac{g_4}2v\pi R\right)\!
	 \cos(\varphi_n-\alpha_n)\cdot\hat\psi_{3L}^{(n)}(x)
	 = 0\ .
\end{align} 

\item
right-handed part
\begin{align}
	\label{piRlhsr2}
  	\cos\!\left(\frac{g_4}2v\pi R\right)\!\sin\varphi_n\cdot\hat\psi_{3R}^{(n)}(x)
	-i\sin\!\left(\frac{g_4}2v\pi R\right)\!
	 \cos(\varphi_n+\alpha_n)\cdot\hat\psi_{2R}^{(n)}(x)
	= 0\ .
\end{align}

\end{itemize}
First, we focus on the left-handed part.
Eqs.~\eqref{piRlhs} and \eqref{piRlhs2} are written as follows;
\begin{align}
	\label{piRlhs3}
	\left[\begin{array}{ccc}
	  \dis-i\sin\!\left(\frac{g_4}2v\pi R\right)\!\cos(\varphi_n+\alpha_n)
	& \dis\cos\!\left(\frac{g_4}2v\pi R\right)\!\sin\varphi_n\\[10pt]
	  \dis\cos\!\left(\frac{g_4}2v\pi R\right)\!\sin\varphi_n
	& \dis i\sin\!\left(\frac{g_4}2v\pi R\right)\!\cos(\varphi_n-\alpha_n)
	\end{array}\right]\!\!\!
	\left[\begin{array}{ccc}
	 \hat\psi_{2L}^{(n)}(x)\\[10pt]
	 \hat\psi_{3L}^{(n)}(x)
	\end{array}\right]
  = \bs0\ .
\end{align}
For \eqref{piRlhs3} to have a non-trivial solution the determinant of the matrix should vanish:
\begin{align}
	\label{nontrivialsol}
	\sin^2\!\left(\frac{g_4}2v\pi R\right)\!\cos(\varphi_n+\alpha_n)\cos(\varphi_n-\alpha_n)
	-\cos^2\!\left(\frac{g_4}2v\pi R\right)\!\sin^2\varphi_n
  = 0\ .
\end{align}
Solving for $\sin^2\!\left(\frac{g_4}2v\pi R\right)$, we get
\begin{align}
	\label{singvpiR}
	\sin^2\!\left(\frac{g_4}2v\pi R\right)\!
  = \frac{m_n^2}{m_n^2-M^2}\sin^2\!\left(\sqrt{m_n^2-M^2}\pi R\right)\ .
\end{align}
Eq.\,\eqref{singvpiR} is an equation to determine the mass eigenvalues $m_n$.
An important fact is there exist two kinds of mass eigenvalues for $n \neq 0$, say $m_n^{(\pm)}$, defined by
\begin{align}
	\label{masseival}
	\frac{m_n^{(\pm)}}{\sqrt{m_n^{(\pm)2}-M^2}}
	\sin\!\left(\sqrt{m_n^{(\pm)2}-M^2}\pi R\right)
  = \pm(-1)^n\sin\!\left(\frac{g_4}2v\pi R\right)\!\ .
\end{align}
Note that $m_n^{(\pm)}$ get degenerate for $v = 0$.
{}From now on the mode sum $\sum_n$ for non-zero KK modes denotes the summation over both eigenstates of $m_n^{(\pm)}$.
Eq.~\eqref{masseival} is a reasonable result, since each of $\psi_2$ and $\psi_3$ has 1 massive Dirac fermion for non-zero KK modes.
Though \eqref{masseival} cannot be analytically solved in general, in the specific case of $M=0$ the solution is easily found to be
\begin{align}
	\label{mnforM0case}
	m_n^{(\pm)}
  = \frac nR\pm\frac{g_4}2v\ .
\end{align}

Once mass eigenvalue is fixed, eq.\,\eqref{piRlhs3} is used to relate $\hat\psi_{2L}^{(n)}(x)$ and $\hat\psi_{3L}^{(n)}(x)$.
{}From \eqref{nontrivialsol} we find
\begin{align}
	\cos\!\left(\frac{g_4}2v\pi R\right)\!\sin\varphi_n
  = \pm\sin\!\left(\frac{g_4}2v\pi R\right)\!
	   \sqrt{\cos(\varphi_n+\alpha_n)\cos(\varphi_n-\alpha_n)}\ .
\end{align}
Thus the ratio of the first row elements of the matrix \eqref{piRlhs3} can be rewritten as
\begin{align}
	\label{psiLratio}
  & \frac{\dis-i\sin\!\left(\frac{g_4}2v\pi R\right)\!
		  \cos\!\left(\varphi_n^{(\pm)}+\alpha_n^{(\pm)}\right)}
		 {\dis\cos\!\left(\frac{g_4}2v\pi R\right)\!\sin\varphi_n^{(\pm)}}
 =  \frac{-i\sqrt{\cos\!\left(\varphi_n^{(\pm)}+\alpha_n^{(\pm)}\right)}}
		 {\pm\sqrt{\cos\!\left(\varphi_n^{(\pm)}-\alpha_n^{(\pm)}\right)}} \ ,
\end{align}
where the relative sign has been fixed so that it recovers the relation in the case of $M=0$ \big($\alpha_n^{(\pm)}=0$\big).

One remark is in order here.
Arguments so far are applicable for all KK modes.
In the case of $n = 0$, however, since $m_0^2 < M^2$ the factor $m_n^{(\pm)2}-M^2$ gets negative and $\sqrt{\smash[b]{m_n^{(\pm)2}-M^2}} = i\sqrt{\smash[b]{M^2 - m_n^{(\pm)2}}}$ becomes pure imaginary.
Thus, {\it e.g.}, $\sin\!\Big(\sqrt{\smash[b]{m_n^{(\pm)2}-M^2}}\pi R\Big)$ should be understood as $i \sinh\!\Big(\sqrt{\smash[b]{M^{2} - m_n^{(\pm)2}}}\pi R\Big)$.

{}From \eqref{psiLratio} we conclude
\begin{gather}
	-i\sqrt{\cos\!\left(\varphi_n^{(\pm)}+\alpha_n^{(\pm)}\right)}
	 \hat\psi_{2L}^{(\pm,n)}(x)
	\pm\sqrt{\cos\!\left(\varphi_n^{(\pm)}-\alpha_n^{(\pm)}\right)}
	\hat\psi_{3L}^{(\pm,n)}(x)
  = 0\notag\\
	\label{psiLrelation}
	\lra~\left\{
	\begin{array}{rl}
	 \hat\psi_{2L}^{(\pm,n)}(x) &\!\!\!\!
   = \dis
	 \sqrt{
	 \frac{\cos\!\left(\varphi_n^{(\pm)}-\alpha_n^{(\pm)}\right)}
		  {2\left|\cos\varphi_n^{(\pm)}\cos\alpha_n^{(\pm)}\right|}
		  }
	 \hat\psi_L^{(\pm,n)}(x)\\[24pt]
	 \hat\psi_{3L}^{(\pm,n)}(x) &\!\!\!\!\dis 
   = \pm i
	 \sqrt{
	 \frac{\cos\!\left(\varphi_n^{(\pm)}+\alpha_n^{(\pm)}\right)}
		  {2\left|\cos\varphi_n^{(\pm)}\cos\alpha_n^{(\pm)}\right|}
		  }
	 \hat\psi_L^{(\pm,n)}(x)
	\end{array}\right.
\end{gather}
where $\hat\psi_L^{(\pm,n)}(x)$ denote the left-handed part of physical quark states $\hat\psi^{(\pm,n)}(x)$ (including non-zero KK modes) with definite 4D masses.
The zero mode $\hat\psi^{(0)}(x)$ is nothing but our down quark $d$.

Let us make comment on a specific case of $\sin\!\left(\frac{g_4}2v\pi R\right) = 1$,
namely $x = \frac{g_4}{2}\pi R v=\frac{\pi}{2}$, for vanishing bulk mass $M=0$.
In this case from \eqref{mnforM0case} we realize $m_n^{(+)} = m_{n+1}^{(-)}$.
Namely, there appears a \lq\lq level crossing" between two mass eigenvalues \big(see \autoref{fig:levelcross2} (a)\big).

In the presence of $M$, however, these levels may mix with each another, because of the breaking of translational invariance in the extra space, and the mass degeneracy is expected to be lifted by an amount $\mathcal O(M)$ \big(see \autoref{fig:levelcross2} (b)\big).
In the case of $x = \frac\pi2$, \eqref{nontrivialsol} implies, independently of the value of $M$, 
\begin{align}
	\label{coscoszero}
	\cos(\varphi_n+\alpha_n)\cos(\varphi_n-\alpha_n) = 0\ .
\end{align}
The equation is invariant under $M\to-M$ ($\alpha_n\to-\alpha_n$),
and we anticipate for small $M$ 
\begin{align}
	m_n^{(+)}
  = \left(n+\frac12\right)\!\frac1R-\mathcal O(M)
	\qquad , \qquad
	m_{n+1}^{(-)}
  = \left(n+\frac12\right)\!\frac1R+\mathcal O(M)\ .
\end{align}
In fact for small $M$, up to $\mathcal O(M)$,
\begin{align}
	\left\{\begin{array}{ccccc}
	 \cos(\varphi_n+\alpha_n) = 0  & \lra &\dis 
	 m_n = \left(n+\frac12\right)\!\frac1R-\frac M{\left(n+\frac12\right)\!\pi}\\[10pt]
	 \cos(\varphi_n-\alpha_n) = 0  & \lra &\dis
	 m_n = \left(n+\frac12\right)\!\frac1R+\frac M{\left(n+\frac12\right)\!\pi}
	\end{array}\right..
\end{align}
Thus we conclude 
\begin{align}
	\label{varphiM0case}
	\cos\!\left(\varphi_n^{(+)}+\alpha_n^{(+)}\right) = 0
	\quad , \qquad
	\cos\!\left(\varphi_{n+1}^{(-)}-\alpha_{n+1}^{(-)}\right) = 0\ .
\end{align}
Then, from \eqref{psiLrelation}
we realize an important property in the specific case $x = \frac\pi2$: 
\begin{align} 
\label{selectionrule1}
	\hat\psi_{3L}^{(+,n)}(x) = 0
	\quad , \qquad
	\hat\psi_{2L}^{(-,n+1)}(x) = 0
	\qquad\big(\,n \geq 0\,\big)\ .
\end{align}

Next, we turn to the right-handed part.
Eqs.\,\eqref{piRlhsr} and \eqref{piRlhsr2} are written as follows:
\begin{align}
	\label{piRlhsr3}
	\left[\begin{array}{ccc}
	  \dis i\sin\!\left(\frac{g_4}2v\pi R\right)\!\cos(\varphi_n-\alpha_n)
	& \dis\cos\!\left(\frac{g_4}2v\pi R\right)\!\sin\varphi_n\\[10pt]
	  \dis\cos\!\left(\frac{g_4}2v\pi R\right)\!\sin\varphi_n
	& \dis-i\sin\!\left(\frac{g_4}2v\pi R\right)\!\cos(\varphi_n+\alpha_n)
	\end{array}\right]\!\!\!
	\left[\begin{array}{ccc}
	 \hat\psi_{3R}^{(n)}(x)\\[10pt]
	 \hat\psi_{2R}^{(n)}(x)
	\end{array}\right]
  = \bs0\ .
\end{align}
We readily find that the determinant of the matrix gives exactly the same relation as \eqref{nontrivialsol} and \eqref{singvpiR}.
Similarly to \eqref{psiLratio} we get
\begin{align}
	\frac{\dis\sin\!\left(\frac{g_4}2v\pi R\right)\!
		  \cos\!\left(\varphi_n^{(\pm)}-\alpha_n^{(\pm)}\right)}
		 {\cos\!\left(\frac{g_4}2v\pi R\right)\!\sin\varphi_n^{(\pm)}}
 &= \frac{i\sqrt{\cos\!\left(\varphi_n^{(\pm)}-\alpha_n^{(\pm)}\right)}}
		 {\pm\sqrt{\cos\!\left(\varphi_n^{(\pm)}+\alpha_n^{(\pm)}\right)}}\ .
\end{align}
Thus we can write as
\begin{gather}
	\label{psiRrelation}
	\left\{
	\begin{array}{rl}
	 \hat\psi_{3R}^{(\pm,n)}(x) &\!\!\!\!\dis
   = \sqrt{
	 \frac{\cos\!\left(\varphi_n^{(\pm)}+\alpha_n^{(\pm)}\right)}
		  {2\left|\cos\varphi_n^{(\pm)}\cos\alpha_n^{(\pm)}\right|}
		  }
	 \hat\psi_R^{(\pm,n)}(x)\\[24pt]
	 \hat\psi_{2R}^{(\pm,n)}(x) &\!\!\!\!\dis 
   = \mp i
	 \sqrt{
	 \frac{\cos\!\left(\varphi_n^{(\pm)}-\alpha_n^{(\pm)}\right)}
		  {2\left|\cos\varphi_n^{(\pm)}\cos\alpha_n^{(\pm)}\right|}
		  }
	 \hat\psi_R^{(\pm,n)}(x)
	\end{array}\right.\ .
\end{gather}
Then from \eqref{varphiM0case} we find in the case of $x = \frac\pi2$
\begin{align} 
\label{selectionrule2} 
	\hat\psi_{3R}^{(+,n)}(x) = 0
	\quad , \qquad
	\hat\psi_{2R}^{(-,n+1)}(x) = 0
	\qquad\big(\,n\geq 0\,\big)\ .
\end{align}
Eq.~\eqref{selectionrule1} together with \eqref{selectionrule2} means that in the specific case of $x = \frac\pi2$, both of $\hat{\psi}^{(+,n)}_L(x)$ and $\hat{\psi}^{(+,n)}_R(x)$ exist only in the position of $\hat{\psi}_2$, while both of $\hat{\psi}^{(-,n)}_L(x)$ and $\hat{\psi}^{(-,n)}_R(x)$ exist only in the position of $\hat{\psi}_3$.
This leads to an impressive conclusion that (all) diagonal Yukawa couplings disappear for $x = \frac\pi2$, as we will demonstrate by explicit calculations below.
This is simply because the Yukawa coupling of the Higgs field $h$ takes a form
\begin{align}
	ig_5\bar\psi\gamma^5 H \frac{\sigma_1}2\psi 
 &= ig_4\bar{\hat\psi}\gamma^5 h\frac{\sigma_1}2\hat\psi
\end{align}
where the matrix $\sigma_1$ connects $\hat{\psi}_2$ and $\hat{\psi}_3$.

\subsection{The normalization of mode functions}

As was mentioned earlier, the normalization factor $\frac1{\sqrt{\pi R}}$ in \eqref{omodefunction}, \eqref{lemodefunction} and \eqref{remodefunction} is the one for $v = 0$ and should be corrected.
{}From \eqref{psiLrelation} and \eqref{psiRrelation}, doublet fermion is written in the following form: 
\begin{align}
	\label{hatpsimefull}
	\hat\psi(x,y)
 &= \sum_n\!
	\left[\begin{array}{c}
	 f_{Le}^{(n)}(y)\hat\psi_{2L}^{(n)}(x)+f_o^{(n)}(y)\hat\psi_{2R}^{(n)}(x)\\[5pt]
	 f_o^{(n)}(y)\hat\psi_{3L}^{(n)}(x)+f_{Re}^{(n)}(y)\hat\psi_{3R}^{(n)}(x)
	\end{array}\right]\notag\\
 &= \sum_n\!
	\left[\begin{array}{c}
	 A_n f_{Le}^{(n)}(y)\hat\psi_L^{(n)}(x)
	 \mp iA_n f_o^{(n)}(y)\hat\psi_R^{(n)}(x)\\[5pt]
	 \pm iB_n f_o^{(n)}(y)\hat\psi_L^{(n)}(x)
	 +B_n f_{Re}^{(n)}(y)\hat\psi_R^{(n)}(x)
	\end{array}\right]
\end{align}
where
\begin{align}
	\label{rationAB}
	A_n^{(\pm)}
  \equiv
	\sqrt{
	 \frac{\cos\!\left(\varphi_n^{(\pm)}-\alpha_n^{(\pm)}\right)}
		  {2\left|\cos\varphi_n^{(\pm)}\cos\alpha_n^{(\pm)}\right|}
		  }
	\quad , \qquad
	B_n^{(\pm)}
  \equiv
	\sqrt{
	 \frac{\cos\!\left(\varphi_n^{(\pm)}+\alpha_n^{(\pm)}\right)}
		  {2\left|\cos\varphi_n^{(\pm)}\cos\alpha_n^{(\pm)}\right|}
		  }\ .
\end{align}
After some straightforward but cumbersome calculations we get relations 
\begin{align}
	\int_{-\pi R}^{\pi R}\hspace{-5mm}dy
	\left(\left|A_n f_{Le}^{(n)}\right|^2\!+\left|B_n f_o^{(n)}\right|^2\right)
  = \int_{-\pi R}^{\pi R}\hspace{-5mm}dy
	\left(\left|A_n f_o^{(n)}\right|^2\!+\left|B_n f_{Re}^{(n)}\right|^2\right)
  = \left|1-\frac{\tan\varphi_n^{(\pm)}\sin^2\!\alpha_n^{(\pm)}}{\varphi_n^{(\pm)}}\right|.
\end{align}
By utilizing these relations we realize that the properly normalized mode functions should be  
\begin{subequations}
\begin{align}
	f_o^{(\pm,n)}(y)
 &= \frac1{\sqrt{\pi RN_n^{(\pm)}}}\sin\!\left(\sqrt{m_n^{(\pm)2}-M^2}\,y\right)\ ,\\
	f_{Le}^{(\pm,n)}(y)
 &= -\frac1{\sqrt{\pi RN_n^{(\pm)}}}
	 \cos\!\left(\sqrt{m_n^{(\pm)2}-M^2}\,|y|+\alpha_n^{(\pm)}\right)\ ,\\
	f_{Re}^{(\pm,n)}(y)
 &= \frac1{\sqrt{\pi RN_n^{(\pm)}}}
	\cos\!\left(\sqrt{m_n^{(\pm)2}-M^2}\,|y|-\alpha_n^{(\pm)}\right)\ ,
\end{align}
\end{subequations}
where
\begin{align}
	N_n^{(\pm)}
 &\equiv
	\left|1-\frac{\tan\varphi_n^{(\pm)}\sin^2\!\alpha_n^{(\pm)}}{\varphi_n^{(\pm)}}\right|
\end{align}
is an additional normalization factor, which becomes 1 for $v = 0$.

\section{Anomalous Higgs interactions}

As was mentioned in the introduction, there exist two (at least) superficially contradictory pictures concerning Higgs interaction with the fermion.
One claims that Higgs interaction is non-linear \big(see \eqref{1.11}\big) and another claims that Higgs interaction is linear as in the standard model \big(see \eqref{1.16}\big).
Before calculating the Higgs interaction explicitly, we first show how these two pictures can be reconciled with each another by a generic argument.

In GHU, Higgs interaction with fermion originates from 5D gauge interaction of $A_y$: 
\begin{align} 
\label{3.0} 
  \mathcal L
  \supset
	g_5\bar\Psi\varGamma^MA_M^a\frac{\lambda^a}2\Psi
  \supset
	ig_5\bar\psi\gamma^5H\frac{\sigma_1}2\psi
  = ig_4\bar{\hat\psi}\gamma^5 h\frac{\sigma_1}2\hat\psi , 
\end{align}
where $A_y^{6(0)} = v_5+H(x)$ and $g_5H(x) = g_4h(x)$.
Substituting the KK mode expansion \eqref{hatpsimefull} for $\hat\psi$ and performing $y$-integral, the fermion's bi-linear form of Yukawa coupling can be written in a matrix form as $hM_Y$, which is clearly linear in $h$ and $M_Y$ is a matrix denoting Yukawa couplings in the base of KK tower of physical quark states $\hat{\psi}^{(\pm, n)}$.
In this base, the 4D mass term of fermion should be written as a diagonalized mass matrix $M_m = {\rm diag}\big(m_{0}, m^{(+)}_1, m^{(-)}_1, \cdots\big)$.
Then the matrix to denote the sum of 4D mass term and Yukawa interaction term, {\it i.e.} \eqref{1.16}, is  
\be \label{3.2}   
	M_m - hM_Y\ .
\ee 
We now realize that $m(v+h)$ such as \eqref{1.11} is nothing but the eigenvalue of the 4D mass operator for the zero-mode fermion, where $h$ is regarded as a constant on an equal footing with the VEV $v$.
Namely, it is an eigenvalue of the matrix given in \eqref{3.2}, $M_m - hM_Y$.
It is reasonable to expect that the eigenvalue is generally non-linear function of $h$, even though the matrix itself is linear in $h$.

In this way, two pictures are known not to be contradictory with each another.
Lesson here is that when $m(v+h)$ is non-linear in $h$, the ``Yukawa coupling matrix" $M_Y$ cannot be diagonal and should contain off-diagonal elements.
This is simply because otherwise the whole matrix $M_m - hM_Y$ gets diagonal and the eigenvalue will be linear in $h$.
In particular, we will show below that in the case of $x = \frac\pi2$ all diagonal elements of $M_Y$ disappear and the Yukawa coupling becomes completely off-diagonal\,!
This seems to coincide with vanishing Yukawa coupling \eqref{1.12} for $x = \frac\pi2$ and suggests that the Yukawa coupling given in \eqref{1.4} corresponds to the diagonal element of $M_Y$.

Here the wisdom in the perturbation theory in quantum mechanics is helpful to understand such equivalence.
It says that the deviation of the ``energy eigenvalue" of state $|n\rangle$ by the perturbative Hamiltonian $H'$ is given by $\langle n|H'|n \rangle$.
Treating $hM_Y$ as $H'$, this means that the deviation of $m(v+h)$ from $m(v)$ at the first order of $h$, {\it i.e.} $\frac{dm(v)}{dv}h = f h$ \big(see \eqref{1.4}\big) should be equal to the diagonal element of $hM_Y$ for the relevant mass eigenstate.
If $m(v)$ denotes the mass function for KK zero mode, the Yukawa coupling of the zero mode is given as 
\be 
\label{3.3}
f = \frac{dm(v)}{dv} = (M_{Y})_{00}\ .
\ee 

We now confirm the equivalence of \eqref{3.3} for arbitrary KK modes by directly calculating the Yukawa coupling in two ways.
First method is to calculate $(M_Y)_{nn}$ for KK $n$-mode by the overlap integral of the mode functions of $\psi^{(\pm, n)}_L$ and $\psi^{(\pm, n)}_R$ \big(The mode function of $A_y^{(0)}$ is just a constant\big):
The $y$-integral of the relevant term
\begin{align}
	\label{yc}
 &  i\frac{g_4}2h(x)\!\!\int_{-\pi R}^{\pi R}\hspace{-5mm}dy
	\left\{\bar{\hat\psi}_2(x,y)\gamma^5\psi_3(x,y)+{\rm h.c.}\right\}\notag\\
 &= -i\frac{g_4}2h\sum_{n,m}\!
	\left\{
	\bar{\hat\psi}_L^{(n)}\hat\psi_R^{(m)}\!\!
	\int_{-\pi R}^{\pi R}\hspace{-5mm}dy
	\left(A_n^*B_mf_{Le}^{(n)*}f_{Re}^{(m)}-B_n^*A_mf_o^{(n)*}f_o^{(m)}\right)
	+{\rm h.c.}
	\right\}.
\end{align}
yields  
\be 
\label{3.4} 
(M_Y)_{nm} = 
-i\frac{g_4}{2} 
	\int_{-\pi R}^{\pi R}\hspace{-5mm}dy
	\left(A_n^*B_mf_{Le}^{(n)*}f_{Re}^{(m)}-B_n^*A_mf_o^{(n)*}f_o^{(m)}\right).
\ee
Let us note in the case $x = \frac\pi2$, the diagonal part ($n = m$) of Yukawa coupling \eqref{3.4} vanishes, since either $A_n$ or $B_n$ vanishes because of \eqref{coscoszero} and \eqref{rationAB}.
The necessary integrals in \eqref{3.4} are given as follows:
\begin{subequations}
\begin{align}
	\int_{-\pi R}^{\pi R}\hspace{-5mm}dy\,f_o^{(n)*}f_o^{(m)}
 &= -\frac1{\sqrt{N_nN_m}}\!
	 \left\{
	 \frac{\sin(\varphi_n+\varphi_m)}{\varphi_n+\varphi_m}
	 -\frac{\sin(\varphi_n-\varphi_m)}{\varphi_n-\varphi_m}
	 \right\}.\\
	\int_{-\pi R}^{\pi R}\hspace{-5mm}dy\,f_{Le}^{(n)*}f_{Re}^{(m)}
 &= -\frac1{\sqrt{N_nN_m}}\!
	 \left\{
	 \frac{\sin(\varphi_n+\varphi_m+\alpha_n-\alpha_m)-\sin(\alpha_n-\alpha_m)}
		  {\varphi_n+\varphi_m}
	 \right.\notag\\
 &\hspace{27mm}\left.
	 +\frac{\sin(\varphi_n-\varphi_m+\alpha_n+\alpha_m)-\sin(\alpha_n+\alpha_m)}
		   {\varphi_n-\varphi_m}
	 \right\}.
\end{align}
\end{subequations}

\subsection{The diagonal Yukawa coupling}

We now focus on the diagonal elements ($n=m$) of Yukawa coupling matrix for arbitrary $n$.
Now the $y$-integrals are simplified as
\begin{align}
	A_n^*B_n\!\int_{-\pi R}^{\pi R}\hspace{-5mm}dy
	\left(f_{Le}^{(n)*}f_{Re}^{(n)}-f_o^{(n)*}f_o^{(n)}\right)
  = -\frac{2A_n^*B_n\cos^2\!\alpha_n}{N_n}\ .
\end{align}
Thus, the diagonal Yukawa coupling is given as:
\begin{align}
	\label{diagonalycc}
	(M_{Y})_{nn}
  = ig_4\frac{A_n^*B_n\cos^2\!\alpha_n}{N_n}
  = i\frac{g_4}2\!
	\left|
	\frac{\varphi_n\cos\alpha_n}{\varphi_n\cot\varphi_n-\sin^2\alpha_n}
	\cot\!\left(\frac{g_4}2v\pi R\right)
	\right|.
\end{align}
We now switch to the second method to get the diagonal Yukawa coupling.
Namely we take the first derivative of mass eigenvalue $m_n$ with respect to $v$.
We recall \eqref{singvpiR}:
\begin{align}
	\sin^2\!\left(\sqrt{m_n^2-M^2}\pi R\right)
  = \left(1-\frac{M^2}{m_n^2}\right)\!\sin^2\!\left(\frac{g_4}2v\pi R\right)\ .
\end{align}
Though this equation cannot be solved analytically for $m_{n}$, we still can get $\frac{dm_n}{dv}$ by differentiating the both sides of the equation with respect to $v$:
\begin{align}
	\label{dmndv}
	\frac{dm_n}{dv}
  = \frac{g_4}2
	\frac{\varphi_n\cos\alpha_n}{\varphi_n\cot\varphi_n-\sin^2\!\alpha_n}
	\cot\!\left(\frac{g_4}2v\pi R\right).
\end{align}
We can confirm that this exactly agrees with \eqref{diagonalycc}.

Generally, for light quark states satisfying $m_n\ll M$ (most probably the zero mode), from \eqref{singvpiR} we easily get 
\begin{align}
	m_n^2 \simeq  \frac{M^2}{\sinh^2(\pi RM)}\sin^2\!\left(\frac{g_4}2v\pi R\right)\ .
\end{align}
Thus,
\begin{align} 
\label{3.4'}
	\frac{dm_n}{dv}
  \simeq \frac{m_n}v\frac{g_4}2v\pi R\cot\!\left(\frac{g_4}2v\pi R\right).
\end{align}

\subsection{On the difference of two pictures} 
The above argument on the Yukawa coupling suggests that as long as we restrict our argument to the zero-mode sector, considering only $(M_Y)_{00}$, the two pictures yield identical prediction.
We, however, anticipate that two pictures provide different predictions on some physical processes, since in the non-linear picture based on $m(v+h)$, the Higgs field, which originally is a dynamical field, is regarded as if it were a constant field.

\begin{figure}[!t]
\centering
\includegraphics[bb=0 0 539 261, scale=0.4]{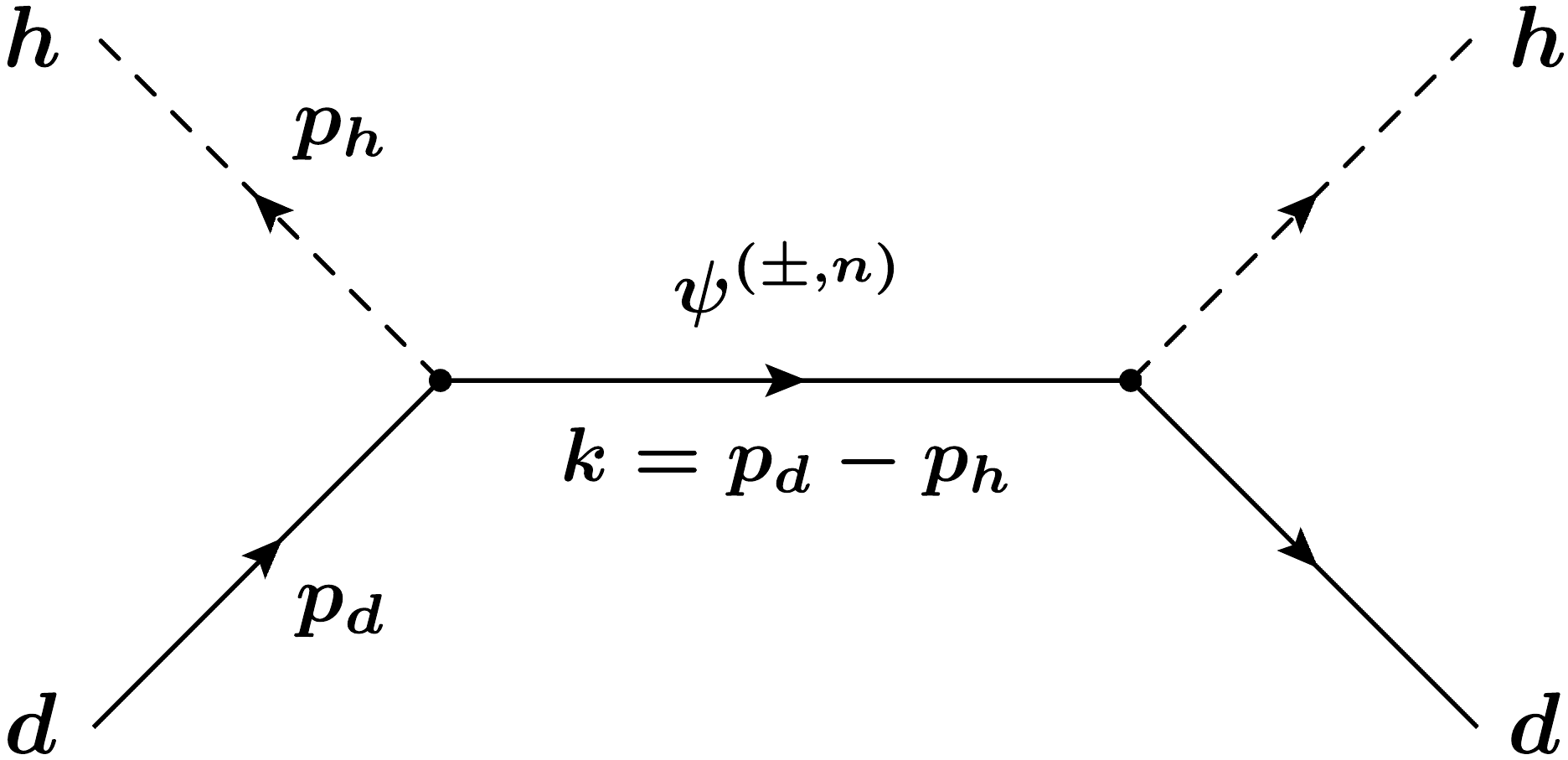}
\caption{Quadratic Higgs interaction with $d$ quark}
\label{fig:Quadratic Interaction}
\end{figure}

In fact it turns out that such difference appears in the quadratic interaction of Higgs.
What we consider is the quadratic Higgs interaction with the zero-mode quark $d$ described by an operator, $\bar d d h^2$.
In the ``linear picture" the quadratic interaction stems from the diagram, as is shown in \autoref{fig:Quadratic Interaction}.
{}From the diagram the Wilson coefficient of the operator $\bar d d h^2$ is calculated to be 
\be 
\label{3.5} 
\sum_{n \neq 0}\frac{|(M_Y)_{n 0}|^2}{k_{\mu}\gamma^\mu - m_n}\ ,
\ee 
where $k_\mu = (p_d)_\mu - (p_h)_\mu$ with $(p_d)_\mu$, $(p_h)_\mu$ being 4-momentum of $d$ quark and the Higgs, respectively.
On the other hand, in the non-linear picture, the quadratic interaction is given by the second derivative of $m_0(v)$ with respect to $v$.
Again the wisdom of quantum mechanics tells us that the second order deviation of energy eigenvalue is given by
\be 
\label{3.6}
 -\sum_{m \neq n}\frac{|\langle m |H'| n \rangle|^2}{E_m-E_n}\ .
\ee 
which means in our case the quadratic term is written as 
\be 
\label{3.7}
m''(v) = - \sum_{n \neq 0}\frac{|(M_{Y})_{n 0}|^{2}}{m_{n}-m_{0}}\ ,
\ee 
since what we are interested in is not energy eigenvalue, but the mass eigenvalue of fermion.
Thus, comparing \eqref{3.5} and \eqref{3.7}, we realize that in the limit where the 4-momentum of the Higgs goes to 0, $(p_h)^\mu \to 0$, $k^\mu$ coincides with $(p_d)^\mu$ and thus by use of the on-shell condition $(p_d)_\mu \gamma^\mu = m_d$ for external  $d$ quark, \eqref{3.5} and \eqref{3.7} are known to just coincide.
This is a reasonable result, since to treat the Higgs field as a constant corresponds to ignoring the 4-momentum of the Higgs.

On the other hand, the above argument implies that in the situation where the Higgs mass and/or Higgs 4-momentum cannot be ignored, {\it i.e.} when Higgs is treated as original dynamical field, as is reasonably expected in the experiments of LHC and linear collider, the two pictures give different predictions on the quadratic Higgs interaction.

\subsection{The deviation of Yukawa coupling from the standard model prediction}
Our formula for the Yukawa coupling of zero-mode fermion is applicable not only for $x = \frac\pi2$, but also for arbitrary $v$.
Thus we now investigate how the Yukawa coupling in GHU deviates from that of the standard model, depending on the VEV $v$ or dimensionless parameter $x$.

{}From \eqref{dmndv}, the diagonal Yukawa coupling of zero-mode fermion in GHU is generally given as 
\begin{align}
	f 
  \equiv
	\frac{dm_0}{dv}\bigg|_{\rm GHU}
  = \frac{g_4}2
	\frac{\varphi_0\cos\alpha_0}{\varphi_0\cot\varphi_0-\sin^2\!\alpha_0}
	\cot\!\left(\frac{g_4}2v\pi R\right).
\end{align}
On the other hand, the Yukawa coupling in the standard model is written as
\begin{align}
	f_{\rm SM} = \frac{m_0}{v}\ .
\end{align}
Therefore, the ratio of these two, indicating the deviation from the standard model prediction when $x \neq 1$, reads as 
\begin{subequations}
\label{dmdvratio}
\begin{align}
	\frac{f}{f_{\rm SM}}
 &= \frac{g_4}2\frac v{m_0}\cdot
	\frac{\varphi_0\cos\alpha_0}{\varphi_0\cot\varphi_0-\sin^2\!\alpha_0}
	\cot\!\left(\frac{g_4}2v\pi R\right)\notag\\
 &= \frac{\bar M^2-\bar m_0^2}
		 {\bar M^2
		 -\bar m_0^2\sqrt{\bar M^2-\bar m_0^2}\coth\!\left(\sqrt{\bar M^2-\bar m_0^2}\right)}
	x\cot x
	\qquad\Big(\,\text{for }M > m_0\,\Big)\\
 &= \frac{\bar m_0^2-\bar M^2}
		 {\bar m_0^2\sqrt{\bar m_0^2-\bar M^2}\cot\!\left(\sqrt{\bar m_0^2-\bar M^2}\right)
		 -\bar M^2}
	x\cot x
	\qquad\Big(\,\text{for }m_0 > M\,\Big)
\end{align}
\end{subequations}
where $\bar M \equiv \pi RM$ and $\bar m_0 \equiv \pi Rm_0$.
For light zero-mode fermion, $m_0 \ll M$, from \eqref{3.4'} we readily know
\begin{align}
	\frac{dm_0}{dv}
  \approx \frac{g_4}2\pi Rm_0\cot\!\left(\frac{g_4}2v\pi R\right).
\end{align}
Therefore, the ratio is approximated as
\begin{align} 
	\frac{f}{f_{\rm SM}}
  \approx \frac{g_4}2v\pi R\cot\!\left(\frac{g_4}2v\pi R\right)
  = x\cot x\ .
\end{align}

Important lesson here is that the deviation disappears in the limit of $x \to 0$ ($x\cot x \to 1$), namely $M_W \ll M_{\rm c} = \frac{1}{R}$ ($M_{\rm c}$ : compactification scale).
This is easy to understand, since in this limit all non-zero KK modes with masses of $\mathcal O(M_{\rm c})$ decouple from the low energy sector: ``decoupling limit".

\subsection{Numerical results}

\begin{figure}[!t]
\centering
\begin{tabular}{ccc}
	\includegraphics[bb=0 0 600 480, scale=0.34]{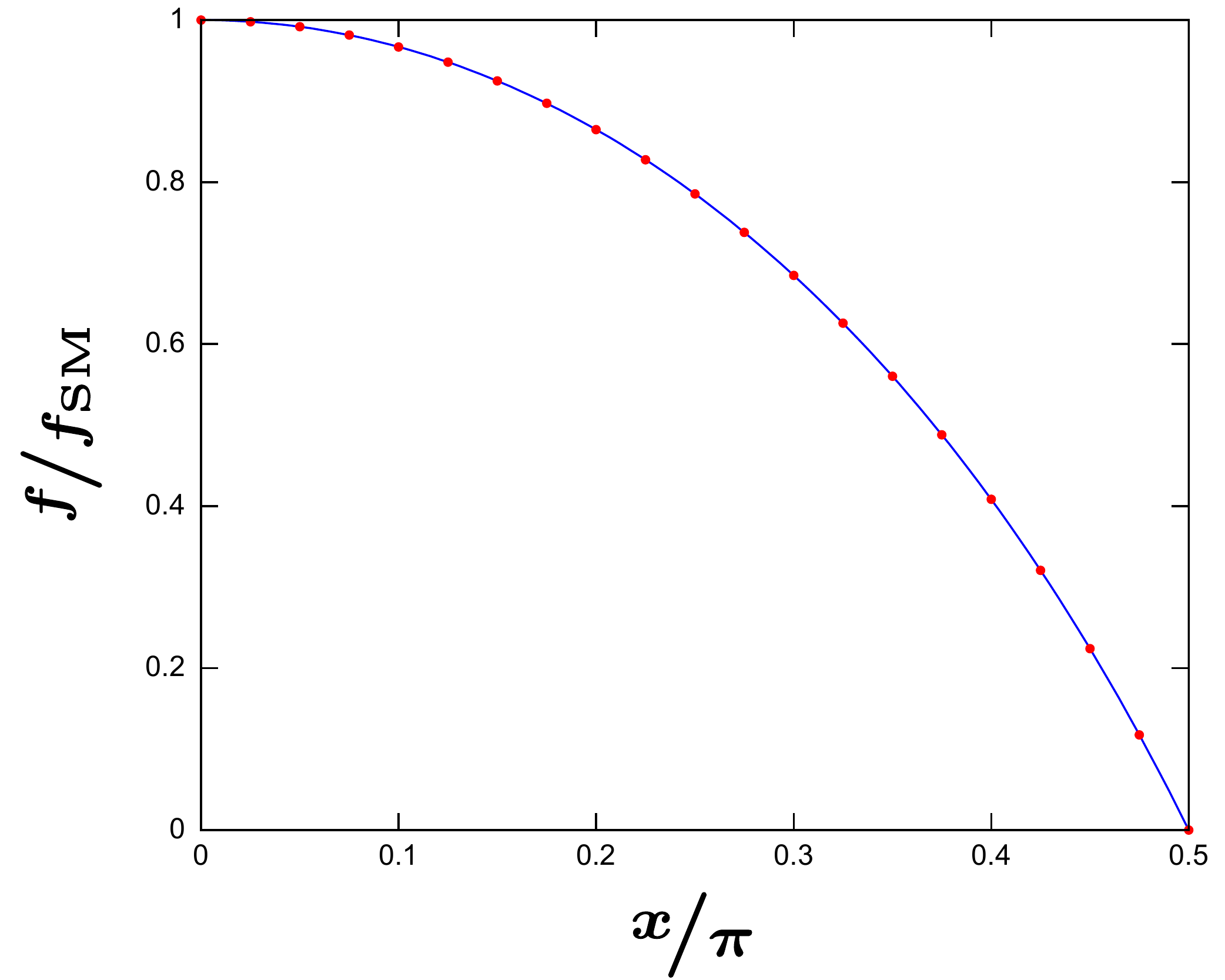}
 && \includegraphics[bb=0 0 600 480, scale=0.34]{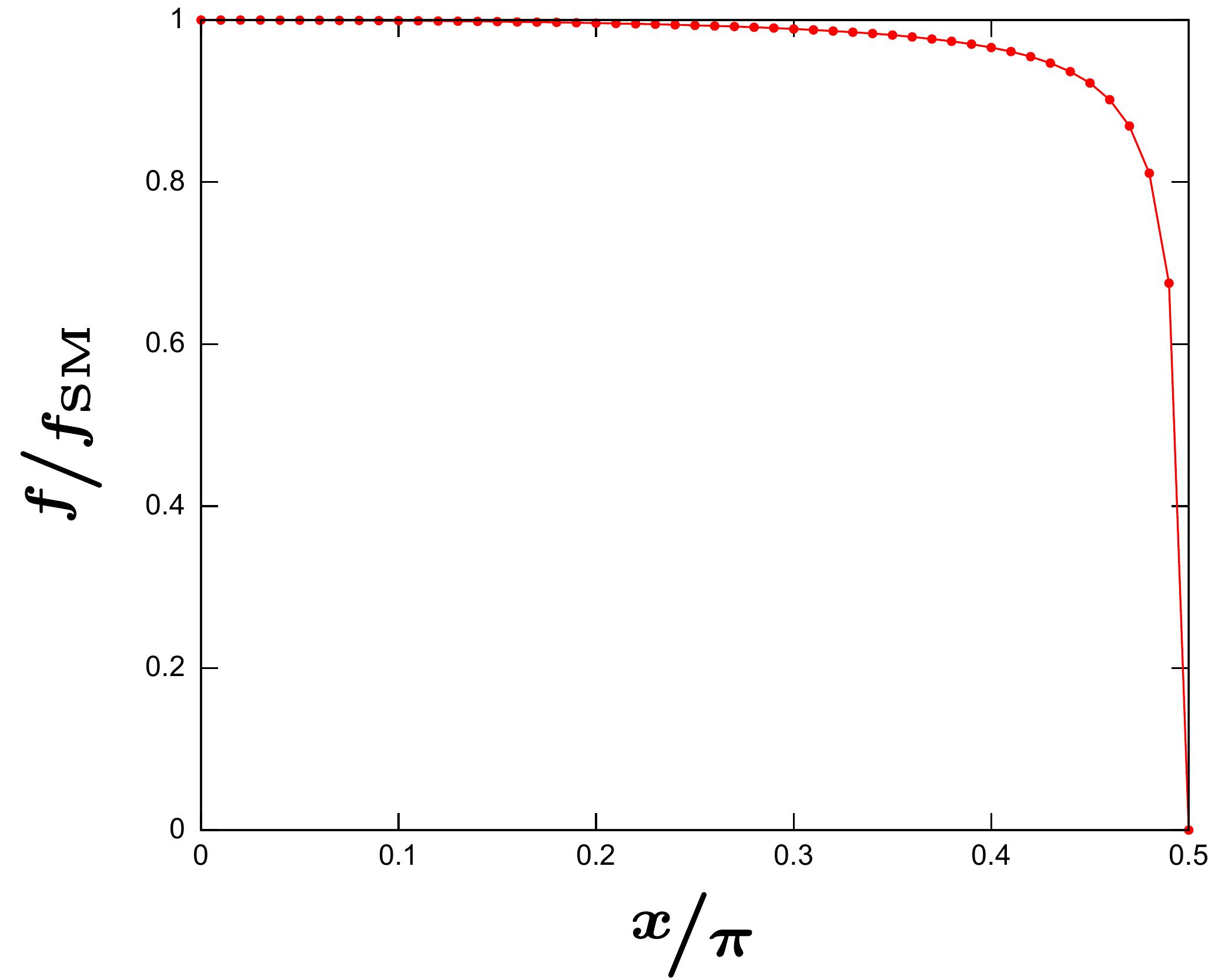}\\[5pt]
	(a)~~$y = \frac{m_b}{M_W} \approx 5.81\times10^{-2}$
 && (b)~~$y = \frac{m_0}{M_W} = 0.99$
\end{tabular}
\caption{The ratio of Yukawa coupling to its standard model prediction for (a) bottom quark and (b) a heavy quark.}
\label{fig:AYukawa}
\end{figure}

In order to calculate \eqref{dmdvratio} numerically, we rewrite this relation in terms of dimensionless parameters $x, y$, and $\bar{M}$ as follows;
\begin{align}
	\frac{f}{f_{\rm SM}}
 &= \frac{x^2y^2-\bar M^2}
		 {x^2y^2\sqrt{x^2y^2-\bar M^2}\cot\!\left(\sqrt{x^2y^2-\bar M^2}\right)-\bar M^2}
	x\cot x
	\quad\left(\,y \equiv \frac{m_0}{M_W} = \frac{\bar m_0}x\,\right).
\end{align}
Once $y$ is fixed by the observed down-type quark mass, the parameter $\bar M$ is determined by $x$ through \eqref{singvpiR} rewritten as
\begin{align}
	\frac{x^2y^2}{x^2y^2-\bar M^2}\sin^2\!\left(\sqrt{x^2y^2-\bar M^2}\right)
  = \sin^2\!x\ .
\end{align}
Thus we can plot the ratio $\frac f{f_{\rm SM}}$ as a function of $x$.
We show some examples in \autoref{fig:AYukawa}.
{\renewcommand{\figureautorefname}{Figure}\autoref{fig:AYukawa}} (a) is for light $b$ quark.
The blue line stands for the function $x\cot x$.
We see that the exact numerical result is very well approximated by an approximated formula $x\cot x$.
On the other hand (b) is for heavy quark, whose mass is comparable with $M_W$.
In this case the bulk mass $M$ should be small and therefore the mass function $m(v)$ approaches to the result of \eqref{mnforM0case}, which is linear in $v$.
Thus in this case the deviation of Yukawa coupling from the standard model prediction is small for broad range of $x$, except the region near to $x = \frac\pi2$, as is seen in the figure.

\begin{figure}[!t]
\centering
\includegraphics[bb=0 0 600 440, scale=0.44]{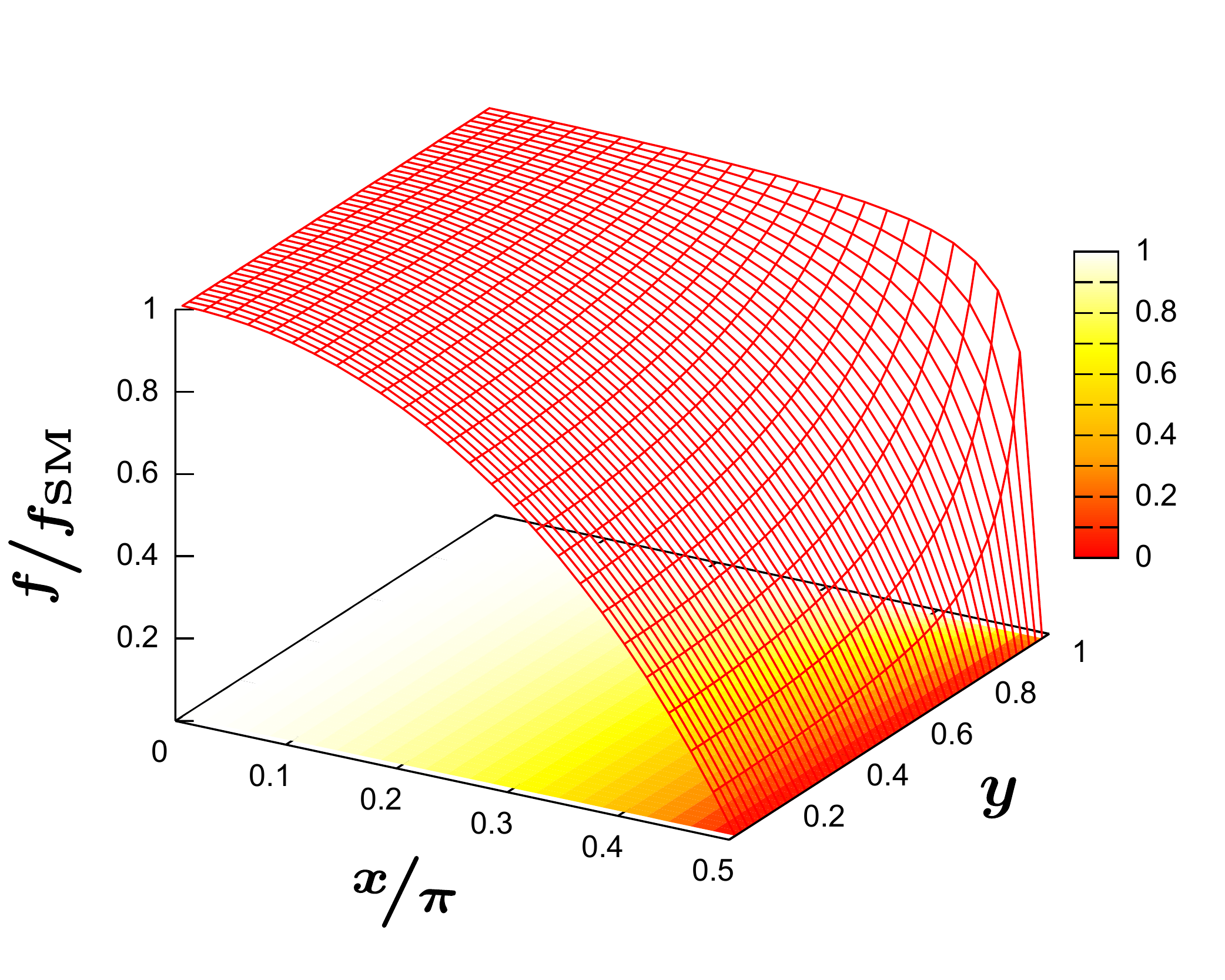}
\caption{The anomalous Yukawa coupling as the function of $\frac x\pi = \frac{g_4}2vR$ and $y = \frac{m_{0}}{M_{W}}$.}
\label{fig:AYukawa3D}
\end{figure}

We have also performed 3D plot of the anomalous Yukawa coupling as a function of $(x,y)$ in \autoref{fig:AYukawa3D}.
As we have already discussed, in the decoupling limit $x \ll 1$ the deviation from the prediction of the standard model disappears, while for relatively large $x$ the deviation becomes remarkable and at the point $x = \frac\pi2$, the Yukawa coupling even vanishes, as was pointed out in refs.\,\cite{2009HK}.

\section{$\bs H$-parity}

The authors of \cite{2009HKT, 2010HTU} have claimed the presence of discrete symmetry, ``$H$-parity", under which only Higgs field, among ordinary standard model fields, changes it's sign.
The parity symmetry exists for the case $x = \frac\pi2$.
Thus the Higgs particle being the lightest among those with odd $H$-parity becomes stable.

Here we would like to formulate $H$-parity from a slightly different point of view in our framework of $SU(3)$ GHU on the flat 5D space-time, and briefly discuss its property.
Basic question we encounter first is whether it is ever meaningful to think about such symmetry.
It stems from the fact that the symmetry seems to be broken anyway by the VEV of Higgs, since Higgs $v + h$ as the whole has odd $H$-parity.%
\footnote{%
Let us note that in \cite{2009HKT, 2010HTU} attitude is a little different and only the physical Higgs field $h$ changes it's sign but not the VEV $v$, though eventually both arguments seem to lead to the same conclusion that $H$-parity exists for $x = \frac\pi2$.
}
Addressing this naive question leads to an important consequence that $H$-parity exists only in the specific situation of $x = \frac\pi2$, as we will see right below.

Let us note that, as is seen in \eqref{3.0}, the Higgs field $h$ connects $\hat{\psi}_2$ and $\hat{\psi}_3$.
It then becomes necessary that $\hat{\psi}_2$ and $\hat{\psi}_3$ have opposite $H$-parities, as long as we impose that the lagrangian is $H$-parity invariant.
On the other hand $H$-parity should be consistent with $SU(2)_L$ symmetry, which remains even after the orbifolding.
Thus the $SU(2)_L$ doublet \big($\hat{\psi}_1$, $\hat{\psi}_2$\big) should have identical $H$-parity.
Thus we are enforced to assign the $H$-parity for the triplet fermion as follows:
\begin{align} 
\label{4.1} 
	H~:\quad\Psi ~\lra~ P\Psi
	\quad , \qquad
	P
  = \left[\begin{array}{ccc}
	 1 & 0 & 0\\[2pt]
	 0 & 1 & 0\\[2pt]
	 0 & 0 & -1\!\!
	\end{array}\right]\ .
\end{align}
Interestingly this transformation is the same as the orbifolding condition \eqref{2.1}, except for the factor $\gamma_5$.
Because of the absence of $\gamma_5$, in contrast to the case of $Z_{2}$ transformation, the Higgs field $v + h$ as the whole changes its sign: $v + h \to -(v + h)$.
Thus at the first glance the non-vanishing VEV $v$ seems to break $H$-parity symmetry spontaneously.
It, however, should be noted that the fact our Higgs field may be physically interpreted as AB phase or Wilson loop makes the situation different.
Namely, in GHU the order parameter for the symmetry breaking should be not the VEV of $A_y$ itself but the VEV of the Wilson loop \big(see \eqref{twistpbc}\big),
\be  
\label{twist'}
\langle W \rangle 
= \e^{i\pi Rg_4v\lambda_6} 
= 
\left[\begin{array}{ccc}
	 1 & 0 & 0\\[2pt]
	 0 & \cos (2x) & i \sin (2x)\\[2pt]
	 0 & i \sin (2x) & \cos (2x)\!\!
	\end{array}\right]. 
\ee 
Because of the presence of the off-diagonal elements, this matrix does not commute with the matrix $P$ in \eqref{4.1}, in general.
For $x \neq 0$ ($v \neq 0$), the only exceptional case is that of $x = \frac\pi2$, where $\sin (2x) = 0$ and the order parameter is clearly $H$-parity invariant:
\be 
\label{4.2} 
P \langle W \rangle P^{-1} = \langle W \rangle\ .
\ee
We thus understand why in the specific case $x = \frac\pi2$ $H$-parity symmetry arises.
Or we may understand the situation by noting that only in the case of $x = \frac\pi2$ \eqref{twist'} is invariant under $v \to -v$ ($x \to -x$).
This reflects the fact that Wilson loop has a periodicity with the period $\pi$ as the function of $x$.
In this way, we can understand why in \cite{2009HKT, 2010HTU} the $H$-parity does not demand the VEV to change its sign.

Let us note that the $H$-parity assignment is the same for both of $\psi$ and $\hat\psi$, since the unitary matrix connecting these fermions appearing in \eqref{hatpsi} is invariant under the transformation:%
\footnote{%
Note that $v$ changes the sign and $\sigma_{1}$ in \eqref{hatpsi} has been replaced by $\lambda_{6}$ as a matrix action on the triplet fermion.
}  
\be 
\label{4.3} 
P \exp\!\Big\{i\frac{g_4}2(-v)y\lambda_6\Big\} P^{-1} 
= \exp\!\Big\{i\frac{g_4}2vy\lambda_6\Big\}\ , 
\ee
which is essentially because $\sigma_{3}(-\sigma_{1})\sigma_{3} = \sigma_{1}$.
At the first glance, \eqref{fermionzeromode} seems to imply that the Higgs is not the lightest particle with odd $H$-parity, since the KK zero mode $d_R$ also seems to have odd parity, belonging to $\psi_3$.
As the matter of fact, however, in the case of $x = \frac\pi2$, {\it i.e.} when $H$-parity exists, both of $d_{R}$ and $d_{L}$ belong to $\hat{\psi}_2$, not to $\hat{\psi}_3$ and therefore have even parity, as was shown in \eqref{selectionrule1} and \eqref{selectionrule2}.
Hence there is no contradiction with the assertion that the Higgs is the lightest $H$-parity odd 
particle.

We also have to confirm that there is no standard model gauge boson having odd $H$-parity.
Once the transformation of triplet \eqref{4.1} is fixed, both of 4D gauge bosons $A_\mu \equiv A^a_\mu\frac{\lambda_a}2$ and 4D scalars $A_y \equiv A^a_y\frac{\lambda_a}2$ have to transform as 
\be 
\label{4.4} 
H~:\quad A_{\mu} ~\lra~ PA_{\mu}P^{-1}
\quad , \qquad
A_{y}~\lra~ PA_{y}P^{-1}\ .
\ee 
Now combining with \eqref{4.1} the lagrangian is clearly $H$-parity invariant.
{}From \eqref{4.4}, all of the standard model gauge bosons belonging to ``unbroken" generators of $SU(2)_L \times U(1)_Y$ are known to have even $H$-parity.
On the contrary, the Higgs belonging to a ``broken" generator has odd $H$-parity, as we expected.
Though would-be NG bosons also have odd parity, they are ``eaten" in the Higgs mechanism by $W^\pm$ and $Z^0$.
Thus we have shown that all standard model particles except for the Higgs have even $H$-parity.
Thus the stability of the Higgs in the case of $x = \frac \pi2$ is shown without going into the details of each interaction.
In fact, from the argument of $H$-parity, we readily know that the diagonal Yukawa coupling of $d$ quark is forbidden for  $x = \frac \pi2$, since such interaction is not $H$-parity invariant.

\section{Higgs interactions with massive gauge bosons} 

We also briefly comment on the Higgs interaction with the massive standard model gauge bosons $W^{\pm}$, $Z^0$.
Concerning the 4D gauge bosons, in contrast to the case of fermions, the mass-squared operator contains only $\partial_{y}$ and the VEV $v$ and therefore the KK mode functions are just ordinary trigonometric functions $\sin (\frac{n}{R}y)$ or $\cos (\frac{n}{R}y)$ depending on the $Z_{2}$ parity.
Thus in this base of mass eigenstates, Higgs interaction terms are also written in the form of diagonal matrices.
Thus mass eigenvalues of the gauge bosons should be linear in $v$, as is shown (for the case of fermion) in \autoref{fig:levelcross2} (a): {\it e.g.} $M_n^{(\pm)} = \frac{n}{R} \pm \frac{g_{4}}{2}v$ $(n \geq 1)$, $M_0 = \frac{g_4}{2}v$ for the case of $W^{\pm}$.
Therefore, the coupling of linear Higgs interactions, $W^{+}_{\mu}W^{- \mu}h$ and $Z^{0}_{\mu}Z^{0 \mu}h$, are readily read off as the derivative of the mass-squared function.
For instance, in the case of $W^{\pm}$, the mass-squared function for the zero mode is $m^2_0(v) = (\frac{g_4v}{2})^2$, and the interaction vertex is given by
\be 
\label{4.5}
i \frac{dm^{2}_{0}(v) }{dv} = i \frac{g_{4}^{2}v}{2} = i g_{4}M_{W},  
\ee
which is the same as the corresponding standard model prediction, in contrast to the result in \cite{2007HS, 2008HOOS} obtained on the curved RS background.
Let us note that in RS space-time, the warp factor clearly breaks translational invariance along the extra space, just as the bulk mass term of fermion on the flat space-time.
This may be an essential reason of such difference of the results concerning Higgs interaction with massive gauge bosons.

However, we now get a little confused, since for the specific case $x = \frac{\pi}{2}$ we have learned from the argument of $H$-parity that the interactions linear in $h$ such as $W^{+}_{\mu}W^{- \mu}h$ are not allowed.
Actually at $x = \frac{\pi}{2}$, as is seen in \autoref{fig:levelcross2} (a), a level crossing occurs and we cannot define the derivative.
We also note that for $x > \frac{\pi}{2}$, as is seen in the figure, the KK zero mode should be replaced by the first KK mode with the mass $m_{1}^{(-)}$.
The derivative of the mass-squared function of this mode $\big\{m_1^{(-)}(v)\big\}^2 = (\frac1R- \frac{g_{4}v}{2})^{2}$ has an opposite sign for that of \eqref{4.5}:
\be 
i\frac d{dv}\left\{m_{1}^{(-)}(v)\right\}^2 = - i g_4\!\left(\frac1R-\frac{g_4v}2\right),
\ee
where $\frac{1}{R}- \frac{g_{4}v}{2}$ is the mass of the lightest $W$ boson for $x > \frac{\pi}{2}$ and should be understood as $M_{W}$.
Thus if there appears a mixing between the zero mode and the first KK mode such level crossing is avoided as is seen in \autoref{fig:levelcross2} (b) and we will be able to claim that the first derivative of mass-squared function can be defined and vanishes for $x = \frac{\pi}{2}$.
We anticipate that the fermion loop causes the mixing because of the fact that the breaking of translational invariance, necessary for the mixing between different KK modes, is realized by the presence of the bulk mass term of fermion.

\section{Summary}

In this paper we discussed the property of Higgs interactions in GHU scenario.
To make the central issue clear, we worked in the simplest framework of the GHU scenario:
$SU(3)$ electroweak model in 5D flat space-time with an orbifold compact space.

In GHU Higgs is identified with the extra space component of higher dimensional gauge field and as the result the Higgs field may be understood as AB phase or Wilson loop when the extra space is compactified on a non-simply-connected space like $S^1$ or $S^1\!/Z_2$.
It was shown to have a very interesting consequence that physical observables are periodic functions of the Higgs VEV $v$.
Correspondingly the Higgs interaction with fermion was argued to be described by non-linear functions of the Higgs field $h$.
This is quite different from the case of the standard model, where Yukawa interaction is of course linear in $h$.
For a specific value of the VEV, the Yukawa coupling is even known to vanish.
Such anomalous Higgs interaction \cite{2007HS, 2008HOOS, 2009HK, 2009HKT, 2010HTU} never appears in the standard model or its straightforward higher dimensional extension, UED theory, and therefore is genuine characteristic prediction of the GHU scenario as a theory of physics beyond the standard model.

By explicit concrete calculation, we have derived formula for the anomalous Yukawa coupling, {\it i.e.} the formula for the deviation of the Yukawa coupling in GHU from that in the standard model as a function of Higgs VEV $v$.
An approximated simple formula for the case of light fermion was also derived.
We have found that for small $x$ ($x = \frac{g_4}{2}\pi R v$), namely when the compactification mass scale $M_{\rm c} = R^{-1}$ is much larger than the weak scale $M_W \sim g_4v$, the deviation is small.
This can be understood as the result of decoupling of massive non-zero KK modes and resultant recovery of the standard model.
On the other hand when $v$ is relatively large the deviation is significant.
And for a specific value of $v$ corresponding to $x = \frac\pi2$, the Yukawa coupling was shown to vanish.

On the other hand, however, even in the GHU model the Yukawa coupling originates from higher dimensional gauge interaction coming from covariant derivative, which is clearly linear in $h$.
We discussed how such apparent contradiction between two ``pictures" (non-linear or linear Higgs interaction with fermion) can be reconciled by utilizing the well-known wisdom of quantum mechanics.
At the same time, we also have demonstrated how these two pictures give different predictions, by taking a typical example of quadratic Higgs interaction.
Such difference turned out to be important when Higgs mass and/or its 4-momentum cannot be neglected, the situation which may be relevant in the experiments of LHC and linear collider.

We also studied ``$H$-parity" symmetry \cite{2009HKT, 2010HTU}, which is utilized to assure the stability of the Higgs field for the specific case $x = \frac\pi2$.
We have identified the operator to fix the eigenvalue of $H$-parity, which turns out to be the same as the matrix $P$ appearing in the $Z_2$ transformation.
At the first glance the existence of Higgs VEV, being odd under $H$-parity transformation, seems to break the $H$-parity symmetry spontaneously.
It, however, was shown that only in the specific case of $x = \frac\pi2$ the VEV of the Wilson loop (the order parameter of symmetry breaking) is invariant under $H$-parity transformation.
This is why in this case $H$-parity is meaningful.
We have confirmed that all standard model particles except for the Higgs have even $H$-parity, as is required to guarantee the stability of the Higgs.

Let us point out one serious problem in the specific case $x = \frac\pi2$.
As the matter of fact, in this case \eqref{twist'} tells us that the VEV of the Wilson loop is a diagonal matrix:
\be  
\label{5.1}
\langle W \rangle 
=  
\left[\begin{array}{ccc}
	 1 & 0 & 0\\[2pt]
	 0 & -1 & 0\\[2pt]
	 0 & 0 & -1\!\!
	\end{array}\right]. 
\ee 
This VEV commutes with two diagonal generators of $SU(3)$.
Hence the rank of the gauge group after the spontaneous symmetry breaking is 2, leaving two $U(1)$ symmetry unbroken, $U(1)_{\rm em}$ and $U(1)_Z$.
The redundant $U(1)_Z$ is associated with $Z^0$ gauge boson and $Z^0$ remains massless even after the spontaneous symmetry breaking, which is not acceptable in order to describe our world.

We have also discussed the Higgs interactions with massive gauge boson, such as $W^{+}_{\mu}W^{- \mu}h$ and $Z^{0}_{\mu}Z^{0 \mu}h$.
In contrast to the case of fermion, the mass eigenvalues are linear in $v$ for gauge bosons and the couplings of the Higgs interactions were shown to be identical with those of the standard model.
Only exception is the case of $x = \frac{\pi}{2}$, where the argument based on the $H$-parity demands that the couplings are absent.
We have argued that in this case there appears a level crossing between the zero mode and the first KK mode, and that once the mixing between these two modes is taken into account the couplings should disappear in accordance with the result based on the $H$-parity.

In the process of the investigation, we encountered some interesting anomalous gauge interactions of massive gauge bosons, $W^\pm$ and $Z^0$, with quarks for relatively large $v$ (and therefore relatively large $x$).
We will report the result of our study concerning the anomalous gauge interactions in a separate paper \cite{2012HKLT}.

\subsection*{Acknowledgments}
It is our great pleasure to thank Y. Hosotani and Y. Sakamura for very useful
discussions and the explanation of their original works.
This work was supported in part by the Grant-in-Aid for Scientific Research of the Ministry of Education, Science and Culture, No.~21244036, No.~23654090, No.~23104009.


\end{document}